\documentclass[a4paper,11pt]{article}
\usepackage{jheppub}
\usepackage{float}
\usepackage{color}
\usepackage{mymacros}
\usepackage{chngcntr}
\usepackage{subcaption}
\usepackage{verbatim}
\usepackage{amsthm}
\usepackage{graphics}
\usepackage{graphicx}
\usepackage{xcolor}
\usepackage{amsmath}
\usepackage{amssymb}
\usepackage{mathrsfs}
\usepackage{bbold}
\usepackage{cases}
\usepackage{epsfig}
\usepackage{epstopdf}
\usepackage{hyperref}
\usepackage{booktabs}

\title{\boldmath Holographic Complexity and Phase Transition for AdS Black Holes}

\author{Hong-Yue Jiang $^{a\,b}$ \footnote{jianghy21@lzu.edu.cn}}
\author{Meng-Ting Wang $^{a\,b}$ \footnote{wangmt21@lzu.edu.cn}}
\author{Yu-Xiao Liu$^{a\,b}$ \footnote{liuyx@lzu.edu.cn, corresponding author}}

\affiliation{
$^{a}$Lanzhou Center for Theoretical Physics, Key Laboratory of Theoretical Physics of Gansu Province, Key Laboratory for Quantum Theory and Applications of the Ministry of Education, Lanzhou University, Lanzhou, Gansu 730000, China\\
$^{b}$Institute of Theoretical Physics $\&$ Research Center of Gravitation, School of Physics Science and Technology Lanzhou University, Lanzhou 730000, China
}

\abstract{Recently, the complexity equals any gravitational observable conjecture has been proposed in [Phys. Rev. Lett. 128, 081602 (2022)], 
which is an extension of the complexity equals volume proposal. 
These gravitational observables are referred to as generalized volumes. 
In this paper, we investigate the generalized volume-complexity for black holes with one or two horizons respectively. 
We verify that the turning time is universal and independent of the Cauchy horizon. 
Not only does this phase transition occur once, but it may also occur two or more times depending on the number and height of the effective potential peaks. 
On the other hand, we confirm that the generalized volume-complexity can be 
divided based on the shape of the effective potential. 
We then discuss the non-smooth transition from the Reissner-Nordström-AdS black hole to the Schwarzschild-AdS black hole. 
}
\begin{document}
\maketitle
\section{Introduction}

In the last three decades, the holographic principle has attracted widespread attention~\cite{Belin:2021bga,Belin:2022xmt,Jorstad:2023kmq,Wang:2023eep,Wang:2023noo,Omidi:2022whq,Baggioli:2021xuv,He:2021xoo,OuYang:2020mpa,Maldacena:1997re,Maldacena:2003nj,Maldacena:2013xja,Susskind:2016jjb,Couch:2016exn,Qu:2022zwq,Qu:2021ius,Shenker:2013pqa,Hartnoll:2008vx,An:2022lvo,Shi:2021rnd,Yang:2019gce,Omidi:2020oit,Akhavan:2019zax,Hartnoll:2020fhc,Stanford:2014jda,Roberts:2014isa,Susskind:2014rva,Susskind:2014moa,Brown:2015bva,Brown:2015lvg,Carmi:2016wjl,Carmi:2017jqz,Chapman:2018dem,Chapman:2018lsv,Swingle:2017zcd,Fu:2018kcp,Cai:2016xho,Yang:2017hbf,Guo:2017rul,Couch:2018phr,Zhang:2022quy,Susskind:2014jwa,Chapman:2016hwi,Cai:2017sjv,An:2018xhv,Goto:2018iay,Bernamonti:2019zyy,Bernamonti:2021jyu}.
In 1997, Maldacena made the first concrete realization of the holographic principle~\cite{Maldacena:1997re}.
The calculation shows that the conformal field theory (CFT) of $d$-dimensional space-time is equivalent to the quantum gravity theory in $(d+1)$-dimensional asymptotic anti-de Sitter (AdS) space-time, known as the AdS/CFT correspondence~\cite{Maldacena:1997re,Maldacena:2003nj}.
After 26 years of development, the AdS/CFT correspondence has become a popular and important research direction in quantum gravity and quantum information.

On the other hand, Susskind and Maldacena proposed a new idea regarding the holographic principle~\cite{Maldacena:2013xja}.
They observed the similarities between the Einstein-Podolsky-Rosen (EPR) paradox and the Einstein-Rosen (ER) bridge and referred to this as the ER = EPR relation~\cite{Maldacena:2013xja,Susskind:2016jjb}.
According to ER = EPR, two systems are connected by an ER bridge if and only if they are entangled.
The ER bridge is not a static object, it grows linearly with time for a very long time.
However, for a black hole system, the dual system reaches thermal equilibrium quickly.
Therefore, entanglement entropy is not enough.
For a complete description of the growth of the ER bridge, Susskind introduced quantum computational complexity as a measure of the volume growth of the ER bridge~\cite{Susskind:2014rva}.

Complexity quantifies the level of difficulty associated with performing a task using a set of simple operations.
In quantum complexity, a quantum circuit is constructed by combining simple gates that act on a few qubits to perform a particular operation using a unitary operator.
In the context of black holes and holography, quantum complexity has recently triggered significant interest, offering a new perspective in the ongoing effort to connect quantum information theory with quantum gravity.
In recent years, there is dramatic progress toward
holographic complexity~\cite{Carmi:2016wjl,Carmi:2017jqz,Chapman:2018dem,Chapman:2018lsv,Swingle:2017zcd,Fu:2018kcp,Cai:2016xho,Yang:2017hbf,Guo:2017rul,Couch:2018phr,Zhang:2022quy,Susskind:2014jwa,Chapman:2016hwi,Cai:2017sjv,An:2018xhv,Goto:2018iay,Bernamonti:2019zyy,Bernamonti:2021jyu,Belin:2022xmt,Belin:2021bga,Jorstad:2023kmq,Wang:2023eep,Omidi:2022whq},
circuit complexity in quantum field theory~\cite{Jefferson:2017sdb,Hackl:2018ptj,Guo:2018kzl,Caceres:2019pgf,Ruan:2020vze},
Krylov complexity~\cite{Parker:2018yvk,Caputa:2021sib,Dymarsky:2021bjq,Avdoshkin:2022xuw,Bhattacharjee:2022vlt,Trigueros:2021rwj,Camargo:2022rnt,Anegawa:2024wov}
and complexity in de Sitter space~\cite{Reynolds:2017lwq,Jorstad:2022mls}.

Under the framework of AdS/CFT, there are three important proposals regarding the holographic complexity:
complexity equals volume (CV)~\cite{Stanford:2014jda,Susskind:2014rva,Susskind:2014moa}, complexity equals action (CA)~\cite{Brown:2015bva,Brown:2015lvg}, and complexity equals spacetime volume (CV 2.0)~\cite{Couch:2016exn}.
The CV proposal maintains that complexity is determined by the maximal volume slice of a hypersurface anchored on the CFT boundary, i.e.
\begin{equation}
    \mathcal{C}_{V}=\max_{\partial\Sigma}\left[\frac{V(\Sigma)}{G_{N}L}\right],
\label{CVp}
\end{equation}
where $\Sigma$ is a hypersurface anchored on the boundary CFT slice, $G_{N}$ denotes Newton's constant in the bulk gravitational theory, and $L$ is a constant with length dimension which often replaced by AdS radius.
Based on the CV proposal, Stanford and Susskind calculated the maximum volume hypersurface inside the AdS black hole after it is disturbed by shock wave geometries and proposed the butterfly effect of complexity~\cite{Stanford:2014jda}. Recently, an analytical expression about the butterfly effect of complexity for an inverted harmonic oscillator was obtained in Refs.~\cite{Qu:2022zwq,Qu:2021ius}.
For the CA proposal, complexity is defined as the action of the Wheeler-de Witt patch, which is given by
\begin{equation}
    \mathcal{C}_{A}=\frac{I_\mathrm{{WdW}}}{\pi \hbar },
\end{equation}
where the Wheeler-de Witt patch is the set of all spacelike hypersurfaces in spacetime~\cite{Carmi:2016wjl}.
And the CV 2.0 proposal is optimized based on the first two conjectures.
In this case, complexity equals the volume of the Wheeler-de Witt patch, that is
\begin{equation}
    \mathcal{C}_{SV}=\frac{V_\mathrm{{WdW}}}{G_{N}L^{2}}.
\end{equation}
Among them, all of holographic complexity proposals have ambiguous definitions, just like the ill-defined constant $L$ in the CV or CV 2.0 proposal,
or a similarly indeterminate length scale that occurs on the null boundaries of the Wheeler-de Witt patch.
In this regard, some scholars believe that this ambiguity is not a shortcoming of the theory itself, but a feature of holographic complexity.
Because there are similar ambiguities in quantum computational complexity, such as the choice of a quantum gate set. The concept of holographic complexity was generalized in Refs.~\cite{Belin:2021bga,Belin:2022xmt,Jorstad:2023kmq}.
The authors defined the new gravitational observable as the generalized volume-complexity, i.e.
\begin{equation}
    \mathcal{O}_{F_{1},\Sigma_{F_{2}}}(\Sigma_\mathrm{{CFT}})=\frac{1}{G_{N}L}\int_{\Sigma_{F_{2}}}d^{d}\sigma\sqrt{h}~F_{1}(g_{\mu\nu};X^{\mu}),
    \label{ef1}
\end{equation}
where $\Sigma_{F_{2}}$ is an codimension-one hypersurface anchored on the CFT slice $\Sigma_\mathrm{{CFT}}$, $h$ is the induced metric on the $\Sigma_{F_{2}}$,
both $F_{1}$ and $F_{2}$ are scalar functions of the bulk metric $g_{\mu\nu}$ and of an embedding coordinate $X^{\mu}$ of the $\Sigma_{F_{2}}$.
If the CFT slice is a constant time slice in the boundary CFT, $\Sigma_\mathrm{{CFT}}$ can be rewritten as $\Sigma_{\tau}$, i.e., $\partial \Sigma_{F_{2}}=\Sigma_{\tau}$.
$\Sigma_{F_{2}}$ is determined by
\begin{equation}
    \delta_{X}\left(\int_{\Sigma_{F_{2}}}d^{d}\sigma\sqrt{h}~F_{2}(g_{\mu\nu}; X^{\mu})\right)=0.
    \label{ef2}
\end{equation}
In general case, there is no correlation between $F_{1}$ and $F_{2}$. However, for $F_{1}=F_{2}=1$, the extremal hypersurface obtained by Eq.~\eqref{ef2} is the extremal volume slice in the CV proposal~\eqref{CVp}.
Extending this example, they analyzed in detail the case where $F_{1}=a_{1}(r),~F_{2}=a_{2}(r)$,
and the conjecture that ``complexity equals anything" is validated across multiple models~\cite{Omidi:2022whq}.

In our previous work~\cite{Wang:2023eep}, we considered the generalized volume-complexity for the four-dimensional Reissner-Nordström-AdS (RN-AdS) black hole, whose growth rates exhibit a discontinues change at a certain boundary time. 
{This discontinuous change in bulk belongs to a phase transition, }
and the transition time is defined as a turning time. 
However, our analysis of the RN-AdS black hole still has certain limitations.
Firstly, It is possible that the existence of the turning time is universal and not unique. 
Therefore, we need to consider more models to get a more general law. 
Secondly, we believe that the impact of the disappearance of the Cauchy horizon on the growth rate of the generalized volume-complexity for black holes with two horizons at late time also deserves discussion.
{Finally, all the above calculations in Ref.~\cite{Wang:2023eep} were done for the case of $F_1 = F_2$. We believe that these questions are also worth exploring for the more general case of $F_1 \neq F_2$.}

In this paper, we would like to investigate the generalized volume-complexity for the planar AdS black hole, the charged BTZ black hole, and the five-dimensional Gauss-Bonnet-AdS black hole with some different gravitational observables. 
And additionally discuss the difference between the RN-AdS black hole at the limit of $Q \to 0$ and $Q = 0$. 
Our findings reveal that the Cauchy horizon is not a necessary condition for the existence of turning time. 
When we choose a suitable gravitational observable, we can always find such a phase transition, whether or not the Cauchy horizon exists. 
Furthermore, there may not be only one turning time, but rather two or even more, depending on the properties of the effective potential. 
On the other hand, we will demonstrate that the generalized volume-complexity can be divided into two categories according to the properties of the effective potential at the singularity or the Cauchy horizon, resulting in different parameter spaces of coupling constants. 
During a supplementary discussion of the RN-AdS black hole, 
we find that with $Q \to 0$, the generalized volume-complexity does not smoothly transform into to the AdS-Schwarzschild solution case.
{In addition, we find that for the more general case of $F_1 \neq F_2$, the existence of the turning time is still universal. 
Different from the $F_1 = F_2$ case, in this situation, the existence of the turning time depends not only on the number of the effective potential peaks decreasing from left to right, but also on the form of $F_{1}$. 
Regarding the growth rate of the generalized volume-complexity, its linear growth in the late-time limit can still be satisfied, although the early evolution is not necessarily monotonic.}

In Sec.~\ref{sec2}, we review the generalized volume-complexity using the planar AdS black hole as an example. 
In Sec.~\ref{sec3}, we analyze the generalized volume-complexity of the black holes with a single horizon. 
Taking the planar AdS black hole and the five-dimensional Gauss-Bonnet-AdS black hole as examples, 
we find that the turning time is independent of the limit of the Cauchy horizon, and divide the generalized volume-complexity into two categories based on the shape of the effective potential. 
In Sec.~\ref{sec4}, we analyse the generalized volume-complexity of the black hole with two horizons. 
Taking the charged BTZ black holes as an example, we examine the generalized volume-complexity constructed by space-time curvature or matter content respectively. 
In addition, we discuss the unsmooth transition of the RN-AdS black hole to the AdS-Schwarzschild black hole. 
{In Sec.~\ref{sec5}, we calculate the generalized volume-complexity and turning time for the $F_1 \neq F_2$ case, and explain how it differs from the case of $F_1=F_2$.}
Finally, the discussion and conclusion are given in Sec.~\ref{sec6}.

\section{Complexity Equals Anything Conjecture} \label{sec2}
In this section, we review the ``Complexity equals anything” conjecture using the planar AdS black hole as an example, presented in Ref.~\cite{Belin:2021bga}. 
The metric of the $(d+1)$-dimensional eternal AdS planar black hole can be described in Eddington-Finkelstein coordinates:
    \begin{gather}
    ds^{2}=-f(r)dv^{2}+2dv dr+\frac{r^{2}}{L^{2}}d \vec{x}^{2},
    \end{gather}
where~$ f(r)=\frac{r^{2}}{L^{2}}\left(1-\frac{r_{h}^{d}}{r^{d}}\right)$ and $v=t+r_{*}(r)$ with $r_{*}(r)=-\int_{r}^{\infty}\frac{dr'}{f(r')}$.
This geometry is regarded as the dual of two decoupled CFTs on planar spatial slices $\Sigma$, which are entangled in the thermofield double (TFD) state
\begin{equation}
    \left| \psi_{\mathrm{TFD}}(\tau)\right \rangle=\sum_{{n}} e^{-\beta E_{n} / 2-i E_{n} \tau}|n\rangle_{\mathrm{L}} \otimes|n\rangle_{\mathrm{R}},
\end{equation}
where $\beta$ is the inverse of the temperature, $E_{n}$ is the energy eigenstate, $L$ and $R$ symbolize the two entangled sides of the planar AdS black hole and label the quantum states $|n\rangle_{L}$ and $|n\rangle_{R}$, respectively.
The time $\tau$ is associated with the left and right boundaries with $t_{L}=t_{R}={\tau}/2$.
According to Ref.~\cite{Belin:2021bga}, when we select $F_{1}=F_{2}=a(r)$, the general expression of complexity can be obtained by
\begin{equation}
    \label{e11}
    \mathcal{C}=\max _{\partial\Sigma=\Sigma_{\tau}}
    \left[\frac{1}{G_{\mathrm{N}} L} \int_{\Sigma} d^{d} \sigma \sqrt{h}~a(r)\right],
\end{equation}
where $\sigma$ is a radial coordinate on the hypersurface $\Sigma$.
Thanks to the planar symmetric, we can parameterize Eq.~\eqref{e11} as
\begin{equation}
    \mathcal{C}=\frac{V_{d-1}}{G_{N}L}\int_{\Sigma}d\sigma \left(\frac{r}{L}\right)^{d-1}\sqrt{-f(r)\dot{v}^{2}+2\dot{v}\dot{r} }~a(r)\equiv \frac{V_{d-1}}{G_{N}L}\int_{\Sigma}d\sigma \mathcal{L}(r,\dot{v},\dot{r}),
    \label{CL}
\end{equation}
where $V_{d-1}$ denotes the volume element of the spatial directions ${\vec{x}}$.
We can refer to $\mathcal{C}$ as the generalized volume-complexity, and treat the integrand as the Lagrangian $\mathcal{L}$.
Choosing the gauge as
\begin{equation}
    \sqrt{-f(r)\dot{v}^{2}+2\dot{v}\dot{r}}= a(r) \left(\frac{r}{L}\right)^{d-1},
\label{gauge}
\end{equation}
we can note that the $\mathcal{L}$ does not depend explicitly on $v$,
so we can define a conserved momentum $P_{v}$:
\begin{equation}
    P_{v}=\frac{a(r) (r/L)^{d-1}(\dot{r}-f(r)\dot{v})}{\sqrt{-f(r)\dot{v}^{2}+2\dot{v}\dot{r}}}=\dot{r}-f(r)\dot{v}.
\label{pv}
\end{equation}
According to Eqs.~\eqref{gauge} and \eqref{pv}, we can obtain the extremality conditions
\begin{align}
    \dot{r}&=\pm \sqrt{P_{v}^{2}+f(r)a^{2}(r)\left(\frac{r}{L}\right)^{2(d-1)}},\\
    \dot{v}&=\frac{1}{f(r)}\left(-P_{v}\pm \sqrt{P_{v}^{2}+f(r)a^{2}(r)\left(\frac{r}{L}\right)^{2(d-1)}} \right).
\end{align}
Therefore, we can explain this problem as the motion of a classical particle in a potential
\begin{equation}
    \dot{r}^{2}+U(r)=P_{v}^{2} ,
    \label{e211}
\end{equation}
where the effect potential $U(r)$ is given by
\begin{equation}
    U(r)=-f(r)a^{2}(r)\left(\frac{r}{L}\right)^{2(d-1)}.
    \label{Ur}
\end{equation}
\begin{figure}[htbp]
    \centering
    \includegraphics[scale=0.12]{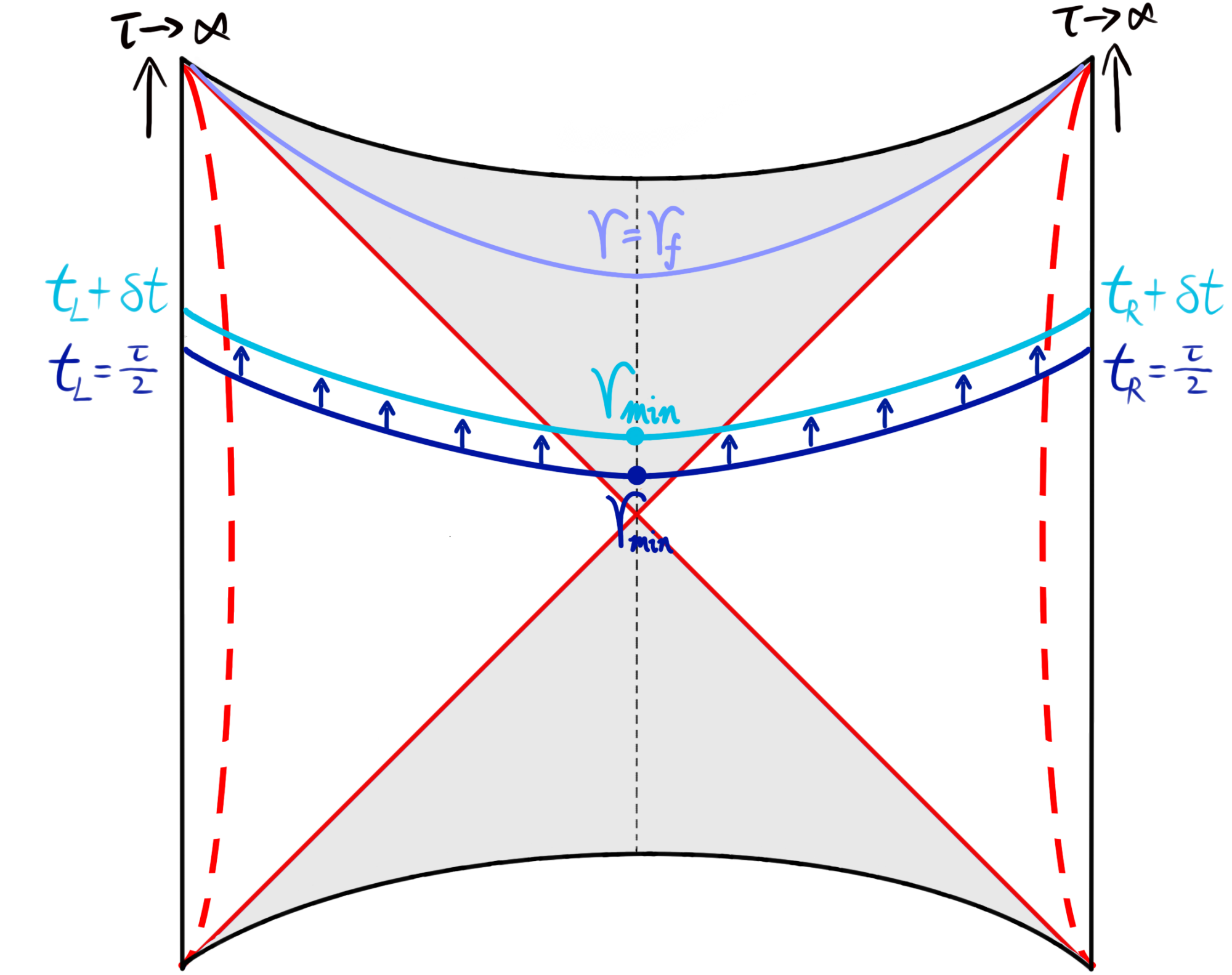}
    \caption[j=2]{The penrose diagram of the eternal AdS black hole. The curves are the extremal hypersurfaces anchored at $t_{L}=t_{R}=\tau/2$. On the black dashed line $t=0$, we have $\dot{r}|_{r=r_{min}}=0$. When we anchor the boundary time as $\tau\rightarrow\infty$, the extremal hypersurface becomes a constant-$r$ slice, i.e. $r=r_{f}$. 
    This image refers to the FIG.1 in Ref.~\cite{Belin:2021bga}.}
    \label{pd1}
\end{figure}
Based on Eq. \eqref{gauge}, we can rewrite Eq. \eqref{CL} as
\begin{equation}
    \mathcal{C}=\frac{V_{d-1}}{G_{N}L}\int_{\Sigma}\frac{ a^{2}(r)\left(\frac{r}{L}\right)^{2(d-1)}}{\dot{r} }dr.
\end{equation}
Using Eq.~\eqref{e211}, we can get $\dot{r}=\pm\sqrt{P_{v}^{2}-U(r)}$. 
On the other hand, we have $U(r_{min})=P_{v}^{2}$, where $r_{min}$ is the minimum radius for the largest generalized volume slice as shown in Fig.~\ref{pd1}. Therefore, $\dot{r}=\sqrt{U(r_{min})-U(r)}$, and we have
\begin{align}
    \mathcal{C}=&\frac{2V_{d-1}}{G_{N}L}\int_{r_{min}}^{r_{\infty}}\frac{a^{2}(r)\left(\frac{r}{L}\right)^{2(d-1)}}{\sqrt{P_{v}^{2}-U(r)}}dr\notag\\
               =&-\frac{2V_{d-1}}{G_{N}L}\int_{r_{min}}^{r_{\infty}}\frac{U(r)}{f(r)\sqrt{U(r_{min})-U(r)}}dr.
\label{Cdr}
\end{align}
On the other hand, according to the ingoing coordinates, we can have
\begin{align}
    \frac{\tau}{2}+r_{*}(r_{\infty})-r_{*}(r_{min})&=\int_{v_{min}}^{v_{\infty}}dv=\int_{r_{min}}^{r_{\infty}}\frac{\dot{v}}{\dot{r}}dr\notag\\
    &=\int_{r_{min}}^{r_{\infty}}dr \left[\frac{1}{f(r)}-\frac{P_{v}}{f(r)\sqrt{P_{v}^{2}+f(r)a^{2}(r)\left(\frac{r}{L}\right)^{2(d-1)}}}\right].
\label{trr}
\end{align}
We hope to extract the part containing Eq.~\eqref{trr} from Eq.~\eqref{Cdr}. By following the method provided in Ref.~\cite{Carmi:2017jqz}, we can rewrite Eq.~\eqref{Cdr} as follows:
\begin{align}
    \mathcal{C}=\frac{2V_{d-1}}{G_{N}L}\int_{r_{min}}^{r_{\infty}}dr\Bigg[-&\frac{P_{v}^{2}}{f(r)\sqrt{P_{v}^{2}+f(r)a^{2}(r)\left(\frac{r}{L}\right)^{2(d-1)}}}\notag\\
    +&\frac{\sqrt{P_{v}^{2}+f(r)a^{2}(r)\left(\frac{r}{L}\right)^{2(d-1)}}}{f(r)}
    +\frac{P_{v}}{f(r)}-\frac{P_{v}}{f(r)}\Bigg].
\end{align}
Therefore, we have
\begin{align}
    \frac{G_{N}L}{2V_{d-1}}\mathcal{C}=&\int_{r_{min}}^{r_{\infty}}dr\left[\frac{\sqrt{P_{v}^{2}+f(r)a^{2}(r)\left(\frac{r}{L}\right)^{2(d-1)}}}{f(r)}-\frac{P_{v}}{f(r)}\right]\notag\\
    &+P_{v}\left[\frac{\tau}{2}+r_{*}(r_{\infty})-r_{*}(r_{min})\right].
\label{Cdrt}
\end{align}
Next, we can take the time derivative of Eq.~\eqref{Cdrt}:
\begin{align}
    \frac{G_{N}L}{2V_{d-1}}\frac{d\mathcal{C}}{d\tau}=&\frac{dP_{v}}{d\tau}\int_{r_{min}}^{r_{\infty}}dr\left[\frac{P_{v}}{f(r)\sqrt{P_{v}^{2}+f(r)a^{2}(r)\left(\frac{r}{L}\right)^{2(d-1)}}}-\frac{1}{f(r)}\right]\notag\\
    &+\frac{dP_{v}}{d\tau}\left[\frac{\tau}{2}+r_{*}(r_{\infty})-r_{*}(r_{min})\right]+\frac{P_{v}}{2}.
\label{dcddt}
\end{align}
We notice that the same term as in Eq.~\eqref{trr} appears in Eq.~\eqref{dcddt}, which can greatly simplify the equation, that is
\begin{equation}
    \frac{d\mathcal{C}}{d\tau}=\frac{V_{d-1}}{G_{N}L}P_{v}.
    \label{CPv}
\end{equation}
Therefore, the growth rate of the complexity is proportional to the conserved momentum $P_{v}$.
Due to $r_{*}=-\int_{r}^{\infty}\frac{dr'}{f(r')}$, we can get the expression of the boundary time by Eq.~\eqref{trr}:
\begin{align}
    \tau=&-2\int_{r_{min}}^{\infty}dr\frac{P_{v}}{f(r)\sqrt{P_{v}^{2}+f(r)a^{2}(r)\left(\frac{r}{L}\right)^{2(d-1)}}} \notag\\
    =&-2\int_{r_{min}}^{\infty}dr\frac{P_{v}}{f(r)\sqrt{P_{v}^{2}-U(r)}}.
\label{tau}
\end{align}
When $\tau \rightarrow \infty$, we let the $U(r_{f})=P_{\infty}$, so we have
\begin{equation}
    \frac{d\mathcal{C}}{d\tau}\bigg\vert _{\tau \rightarrow \infty}=\frac{V_{d-1}}{G_{N}L}P_{\infty }.
    \label{CPvinfty}
\end{equation}

\section{Holographic Complexity for Black Holes with A Single Horizon}\label{sec3}
In this section, we will take the planar AdS black hole and the five-dimensional Gauss-Bonnet-AdS black hole as examples to discuss the generalized volume-complexity for the black holes with a single horizon and a space-like curvature singularity.
We find that in such case, the turning time still exists and may not be unique. 
On the other hand, for the five-dimensional Gauss-Bonnet-AdS black hole, we find that the shape of the effective potential can determine whether the parameter space of the coupling constant is the full space. 

\subsection{Planar AdS Black Hole}
\begin{figure}[htbp]
    \centering
    \begin{subfigure}[b]{0.45\textwidth}
        \centering
		\includegraphics[scale=0.18]{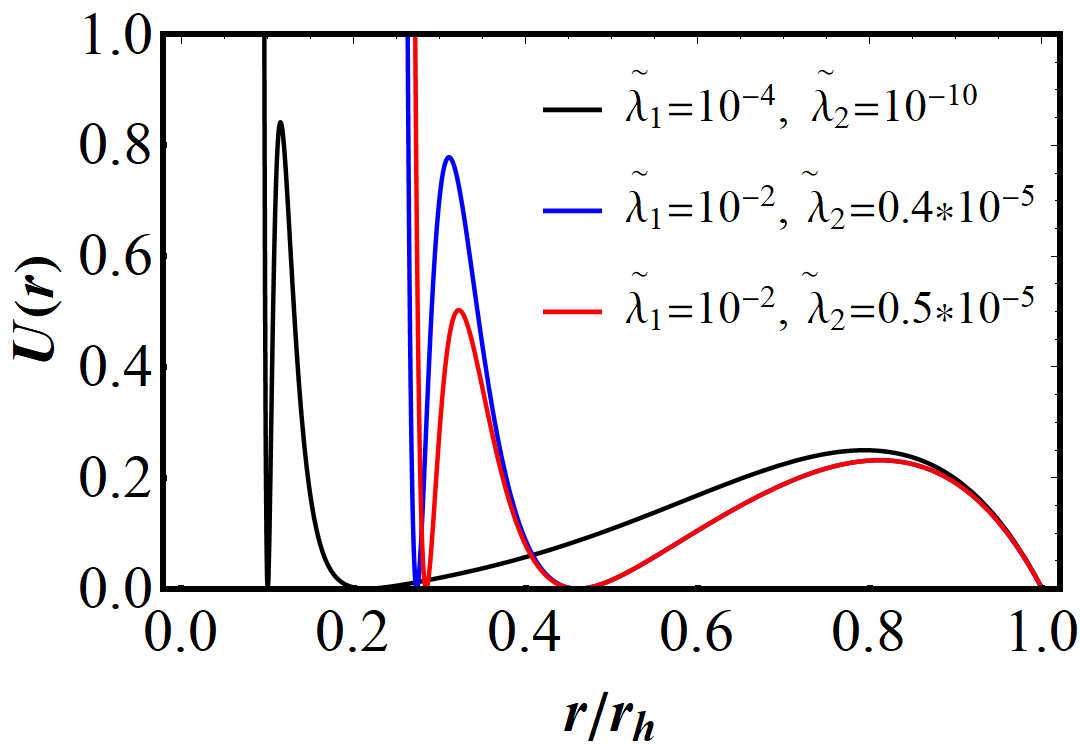}
        \caption{The effective potential defined in Eq.~\eqref{url1l2}.}
        \label{SWXUR2a}
    \end{subfigure}
    \hfill
    \begin{subfigure}[b]{0.45\textwidth}
        \centering
        \includegraphics[scale=0.155]{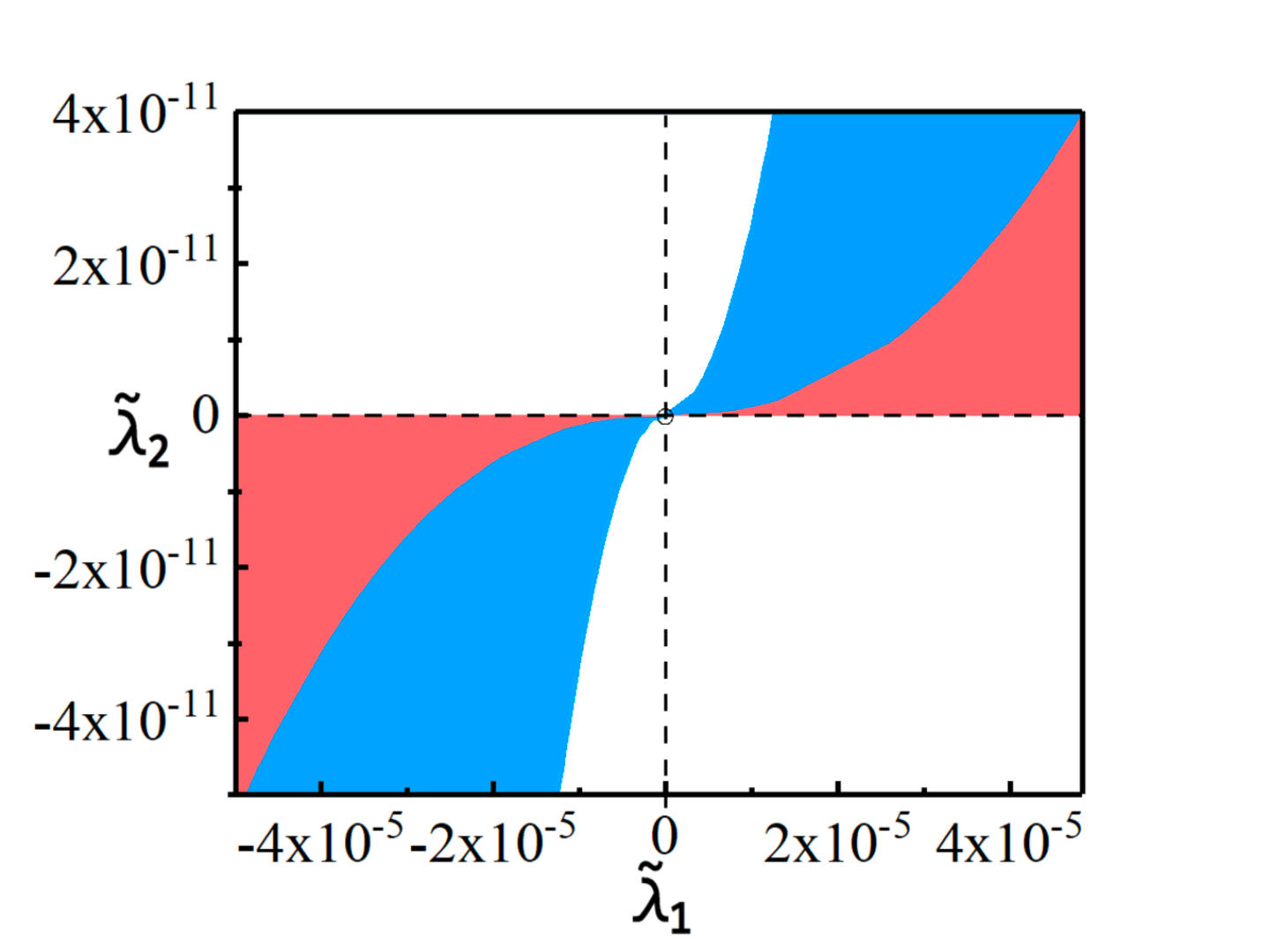}
        \caption{The parameter space of $\tilde{\lambda}_{1}$ and $\tilde{\lambda}_{2}$.}
        \label{SWXUR2b}
    \end{subfigure}
    \caption{{(a) The image of the effective potential $U(r)$ with different coupling constants.
	         (b) In the red area, the effective potential has two local maxima, and the left one is larger than the right one.
	             In the blue area, there are also two local maxima, but the local maxima on the right is greater than the one on the left.
	}}
	\label{SWXUR2}
\end{figure}

\begin{figure}[htbp]
    \centering
    \begin{subfigure}[b]{0.45\textwidth}
        \centering
		\includegraphics[scale=0.15]{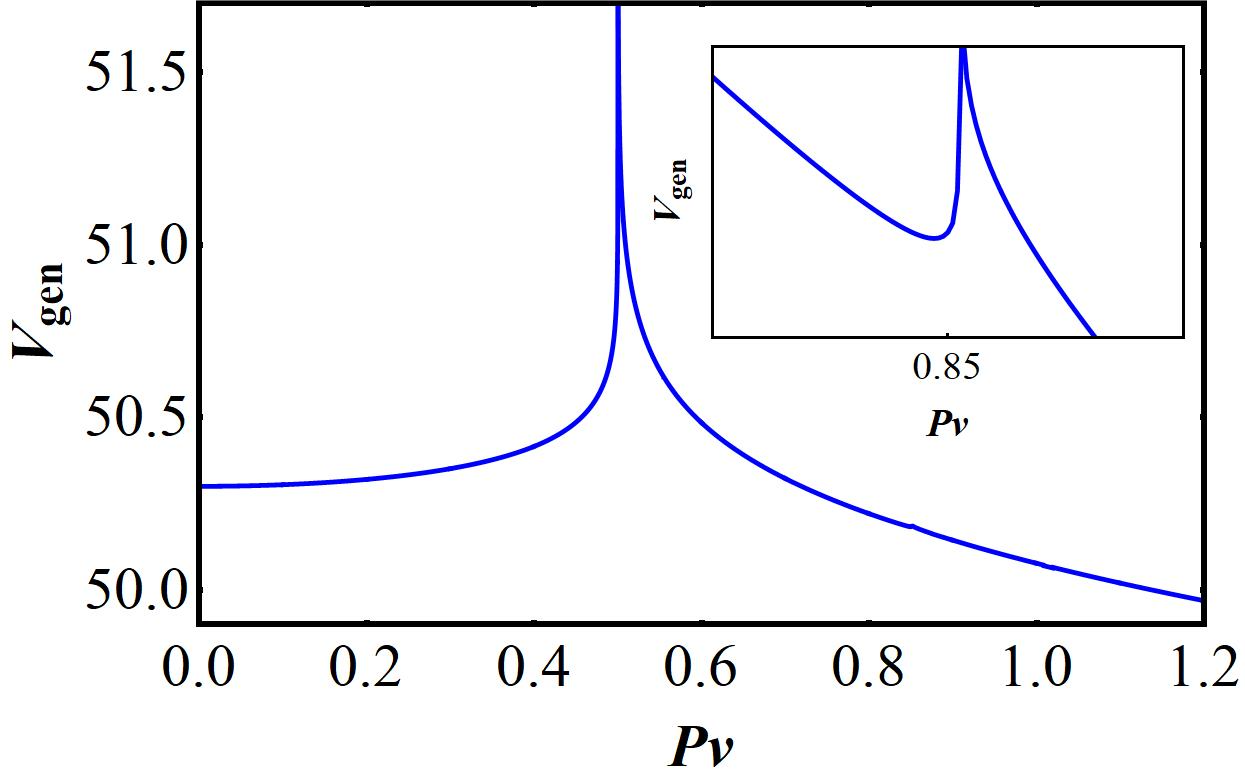}
        \caption{Generalized volume defined in Eq.~\eqref{Vgen}.}
        \label{PvVa}
    \end{subfigure}
    \hspace{0.05\textwidth}
    \begin{subfigure}[b]{0.45\textwidth}
        \centering
        \includegraphics[scale=0.15]{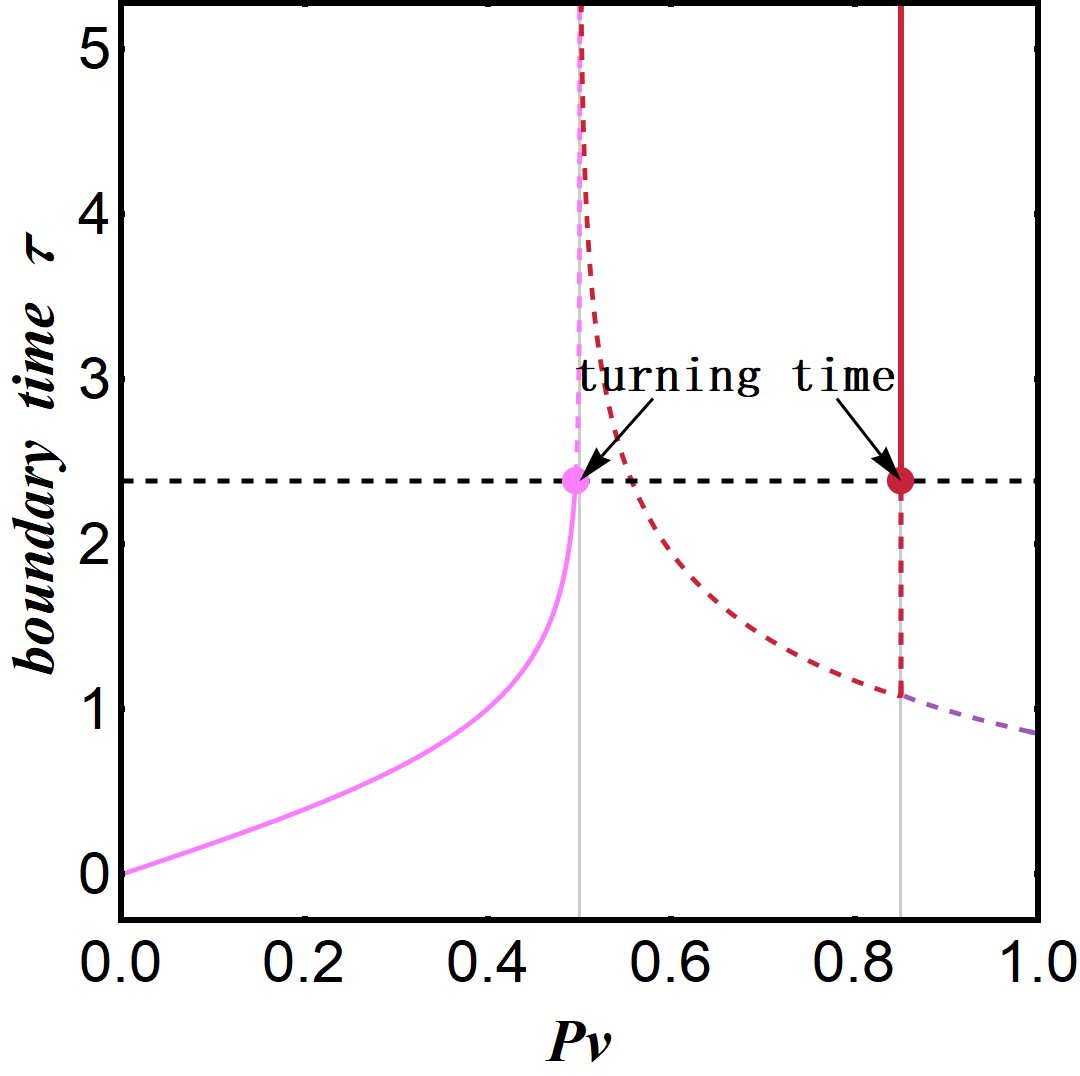}
        \caption{Boundary time $\tau$.}
        \label{PvVb}
    \end{subfigure}
    \caption{(a) The relation between the generalized volume and the conserved momentum $P_{v}$ of the planar AdS black hole with $d=3,~\frac{r_{h}}{L}=1,~\tilde{\lambda}_{1}=10^{-4},~\tilde{\lambda}_{2}=1.1*10^{-10}$.
	The part outside the horizon grows linearly with $r$ and diverges at infinity, we just cutoff it at $r=10r_{h}$.
	(b) The relation between the boundary time $\tau$ and the conserved momentum $P_{v}$ using the parameters mentioned above.
	The two solid grey lines represent the two local maxima of $U(r)$, i.e., $U(r_{f})$.
	The pink curve corresponds to the right peak, the red curve corresponds to the left one, and the purple curve corresponds to the divergent branch on the left. 
	Both the pink and red solid lines represent larger generalized volumes at the corresponding boundary time, while the dotted lines are the opposite. }
		\label{PvV}
\end{figure}

There are some examples of $a(r)$ presented in Ref.~\cite{Jorstad:2023kmq}.
In this section, we expand on their examples and consider the turning time in each other.
We also use the Weyl tensor for the bulk spacetime to construct the gravitational observable on the codimension-one extremal slice $\Sigma$.
That is
\begin{align}
    a(r)=&1+\sum_{j=1}^{N}(-1)^{j}\lambda_{j}(L^{4}C^{2})^{j},\\
    C^{2}\equiv& R_{\mu \nu \sigma \rho}R^{\mu \nu \sigma \rho}-\frac{4}{d-1}R_{\mu \nu}R^{\mu \nu}+\frac{2}{d(d-1)}R=d(d-1)^{2}(d-2)\frac{r_{h}^{2d}}{L^{4}r^{2d}},
\end{align}
where $C^{2}\equiv C_{\mu \nu \sigma \rho}C^{\mu \nu \sigma \rho}$ denotes the square of the Weyl tensor for the bulk spacetime, $\lambda_{j}$ are dimensionless coupling constants, and $j=1,2, \cdots, N$.
To simplify the calculation, we set $\tilde{\lambda}_{j}=[d(d-1)^{2}(d-2)]^{j}\lambda_{j}$.
After this substitution, $a(r)$ can be rewritten as
\begin{align}
    a(r)=1+\sum_{j=1}^{N}(-1)^{j}\tilde{\lambda}_{j}\left(\frac{r_{h}}{r}\right)^{2dj}.
\end{align}
When all $\lambda_j=0$, we have $a(r)=1$, which corresponds to the CV proposal. The case of $N=1$ was discussed in Ref.~\cite{Belin:2021bga}. We begin our analysis with $N=2$, i.e.,
    \begin{align}
        a(r)=1-\tilde{\lambda}_{1}\left(\frac{r_{h}}{r}\right)^{2d}+\tilde{\lambda}_{2}\left(\frac{r_{h}}{r}\right)^{4d}.
    \end{align}
It was chosen in Ref.~\cite{Jorstad:2023kmq} to make the extreme surface approaches the singularity.
Then, we can obtain the expression of the effective potential, i.e.,
    \begin{align}
        U(r)=-r^{-d}\left(\frac{r}{L}\right)^{2d}(r^{d}-r_{h}^{d})
        \left[1-\left(\frac{r_{h}}{r}\right)^{2d}\tilde{\lambda}_{1}
        +\left(\frac{r_{h}}{r}\right)^{4d}\tilde{\lambda}_{2}\right]^{2}.
    \label{url1l2}
    \end{align}
We consider the case of $d=3$ and choose appropriate value of the coupling parameter $\tilde{\lambda}_{j}$ as shown in Fig.~\ref{SWXUR2b}.
In the red area, $U(r)$ gives two peaks, and their values gradually decrease from left to right.
According to Eq.~\eqref{Cdr}, we can obtain the  expression for the generalized volume, which is given by
\begin{equation}
    V_{gen}=V_{d-1}\int_{r_{min}}^{r_\infty }\frac{a^{2}(r)\left(\frac{r}{L}\right)^{2(d-1)}}{\sqrt{Pv^{2}-U(r)}}dr.
\label{Vgen}
\end{equation}
By calculating the generalized volume, we can observe its correspondence with $P_{v}$ as shown in Fig.~\ref{PvVb}.
When we select the appropriate value of $\lambda$ to construct the $U(r)$ as shown in Fig.~\ref{SWXUR2a}, there are two local maxima of the effective potential.
Equation~\eqref{CPv} tells us that a higher peak corresponds to a larger growth rate of the generalized volume-complexity.
Therefore, at a very late time, the generalized volume corresponding to the peak on the left must be bigger than that of the right.
Interestingly, as shown in Fig.~\ref{PvVa}, the shorter peak corresponds to a larger generalized volume over a significant range.
Consequently, there must be a moment when the generalized volumes corresponding to the left and right peaks are equal, and after that, the relative size relationship between these two will reverse. We refer to this moment as the turning time\cite{Wang:2023eep}.

\begin{figure}[htbp]
    \centering
    \begin{subfigure}[b]{0.45\textwidth}
        \centering
		\includegraphics[scale=0.3]{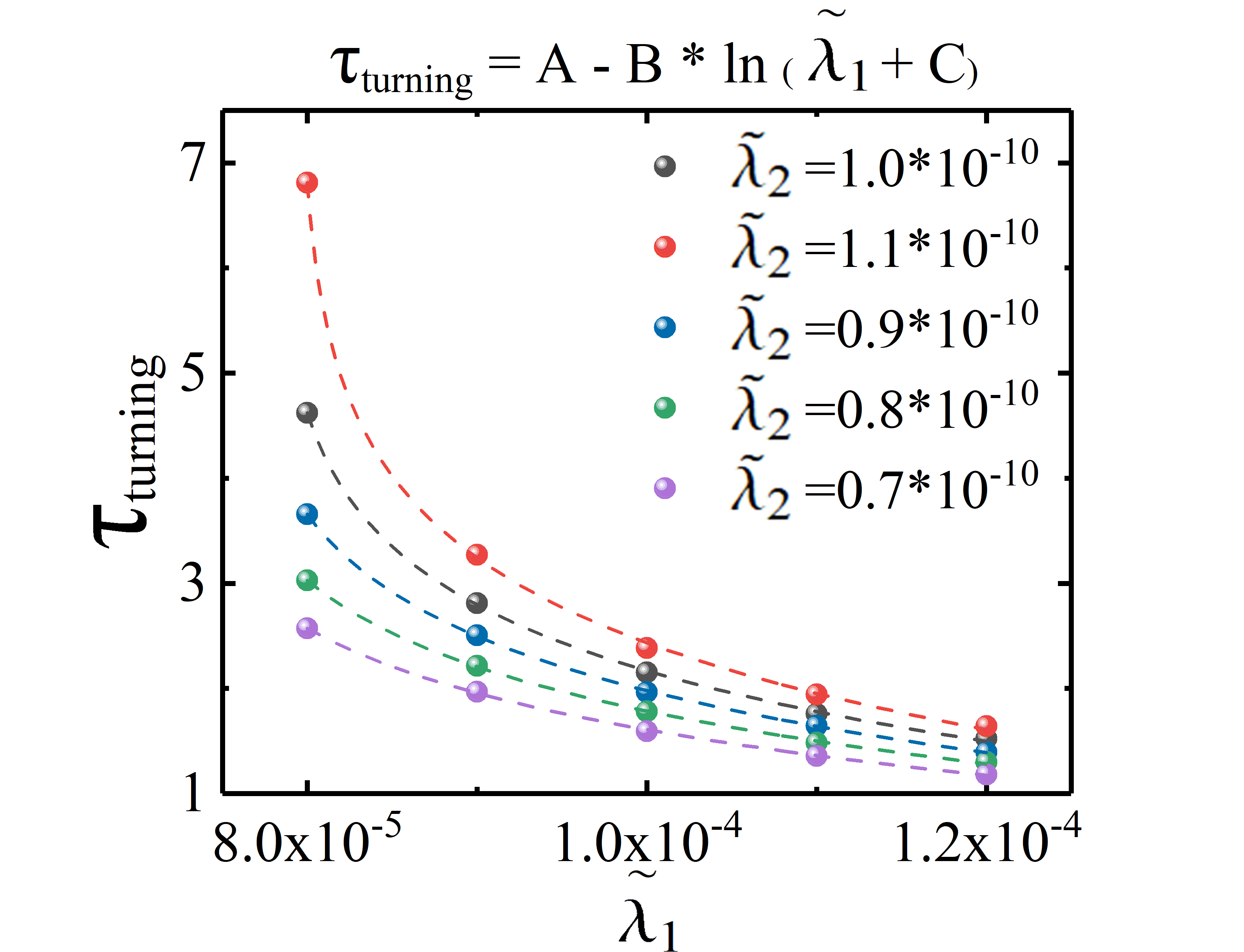}
        \caption{}
        \label{ttswxa}
    \end{subfigure}
	\hspace{0.05\textwidth}
    \begin{subfigure}[b]{0.45\textwidth}
        \centering
        \includegraphics[scale=0.3]{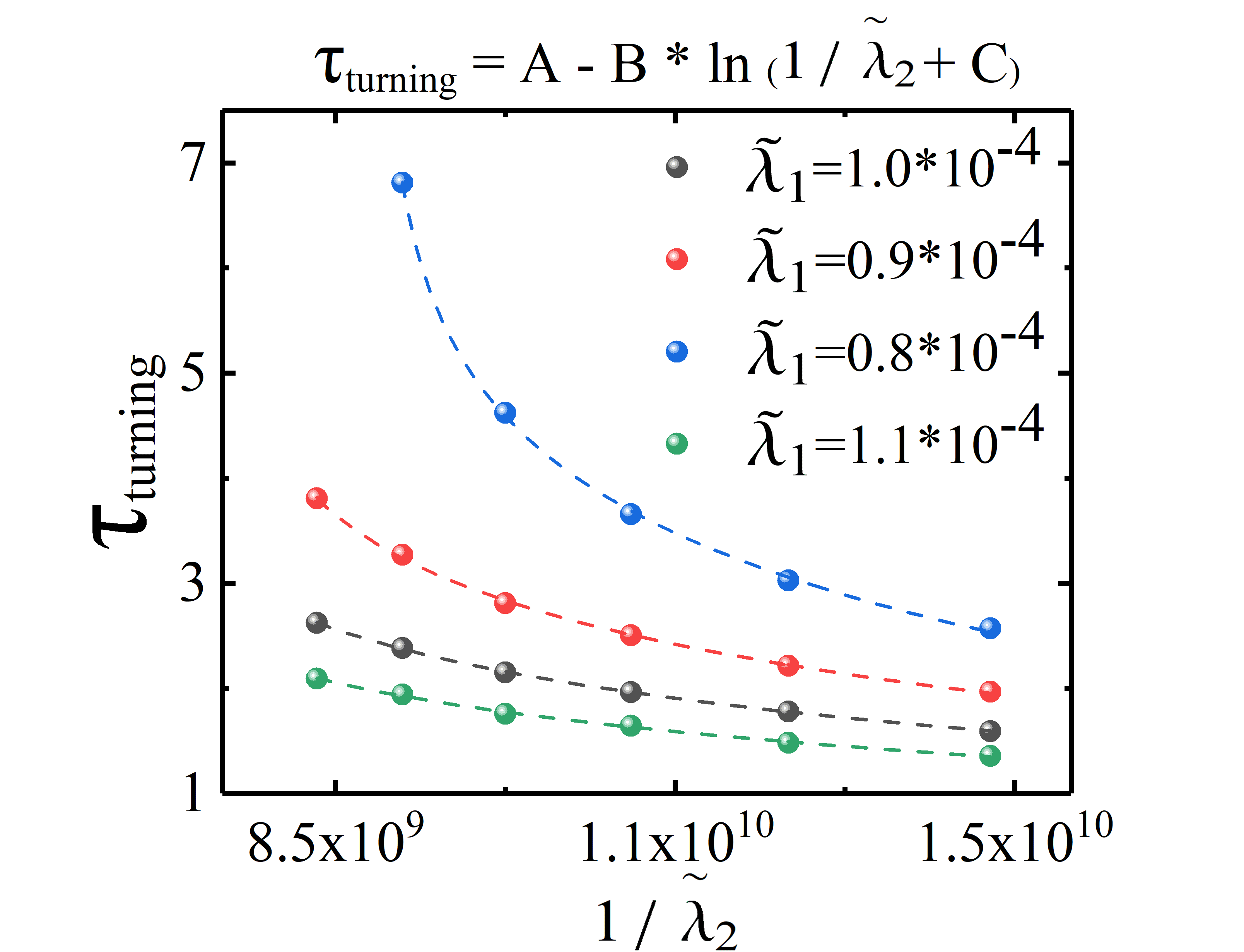}
        \caption{}
        \label{ttswxb}
    \end{subfigure} 
    \caption{The fitting of turning time on $\tilde{\lambda}_{1}$ and $\tilde{\lambda}_{2}$. }
		\label{ttswx}
\end{figure}

By selecting different values of $\tilde{\lambda}_{1}$ and $\tilde{\lambda}_{2}$, we can get the corresponding relationship between the parameters and the turning time, as shown in Fig.~\ref{ttswx}.
In another paper, we investigated the turning time for the four-dimensional RN-AdS black hole~\cite{Wang:2023eep}.

We discoverd that the turning time has a logarithmic dependence on its parameter, and we have determined the fitting function to be given by
\begin{align}
\tau_{turning}=&a_{1}-b_{1}\ln (\tilde{\lambda}_{1}+c_{1}),\\
\tau_{turning}=&a_{2}-b_{2}\ln (1/\tilde{\lambda}_{2}+c_{2}),
\end{align}
where $a_{i}$, $b_{i}$, and $c_{i}$ are the fitting values.
Among them, $a_{i}$ and $b_{i}$ have dimensions of length, while $c_{i}$ are dimensionless constants.

\begin{figure}[htbp]
    \centering
    \begin{subfigure}[b]{0.45\textwidth}
        \centering
		\includegraphics[scale=0.18]{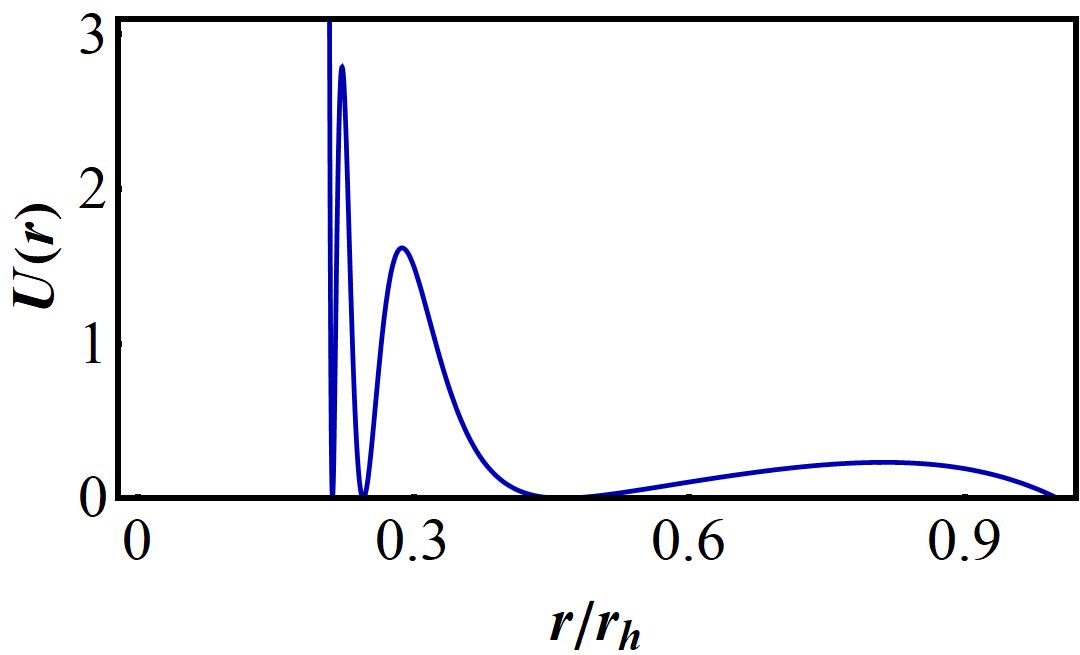}
        \caption{Effective potential $U(r)$.}
        \label{SWXUR3a}
    \end{subfigure}
	\hspace{0.05\textwidth}
    \begin{subfigure}[b]{0.45\textwidth}
        \centering
        \includegraphics[scale=0.17]{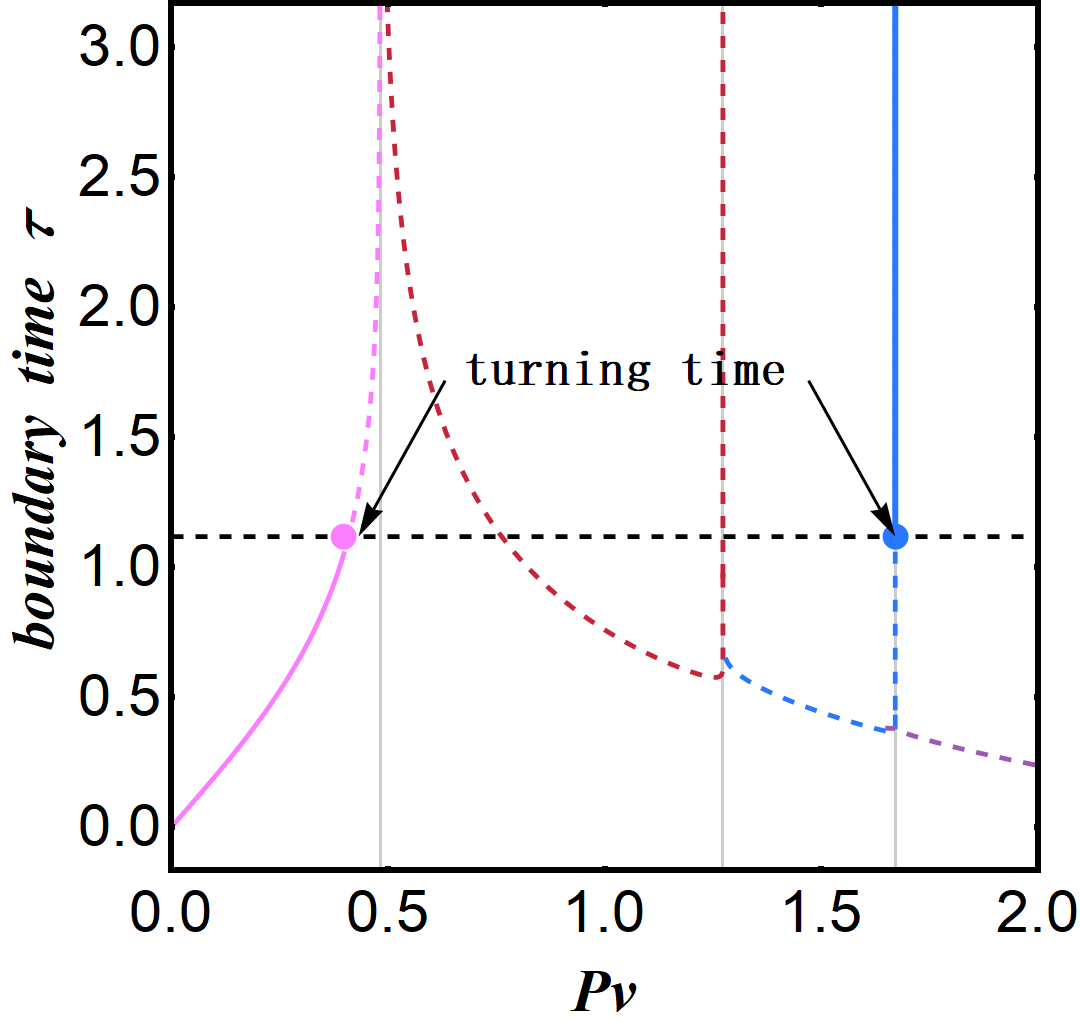}
        \caption{Boundary time $\tau$.}
        \label{SWXUR3b}
    \end{subfigure}
    \caption{{(a) The image of the effective potential $U(r)$ with $L=1$, $\tilde{\lambda}_{1}=10^{-2}$, $\tilde{\lambda}_{2}=3*10^{-6}$ and $\tilde{\lambda}_{3}=1.9*10^{-10}$.
    (b) The relation between the boundary time $\tau$ and the conserved momentum $P_{v}$ using the parameters mentioned above.
    The three solid grey lines represent the three local maxima of $U(r)$, i.e., $U(r_{f})$.
    The pink, red and blue curves correspond to the right, middle and left peaks respectively, and the purple curve corresponds to the divergent branch on the left.
    Both the pink and blue solid lines represent larger generalized volumes at the corresponding boundary time, while the dotted lines are the opposite.} }
		\label{SWXUR3}
\end{figure}

\begin{figure}[htbp]
	\centering
	\begin{subfigure}[b]{0.45\textwidth}
		\centering
		\includegraphics[scale=0.2]{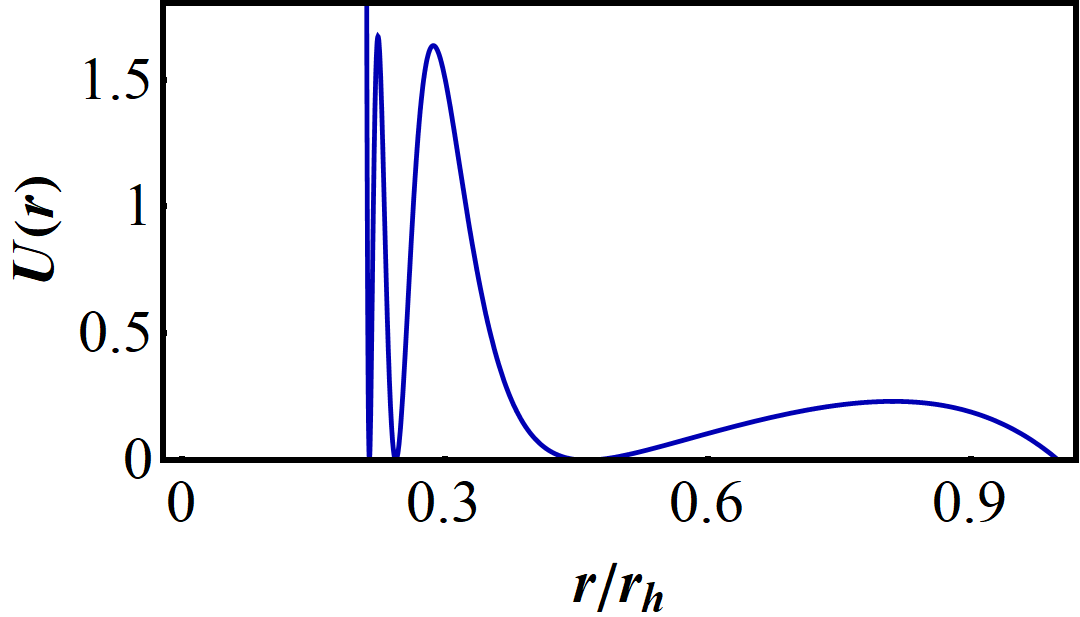}
		\caption{Effective potential $U(r)$.}
		\label{SWXUR32a}
	\end{subfigure}
    \hspace{0.05\textwidth}
	\begin{subfigure}[b]{0.45\textwidth}
		\centering
		\includegraphics[scale=0.17]{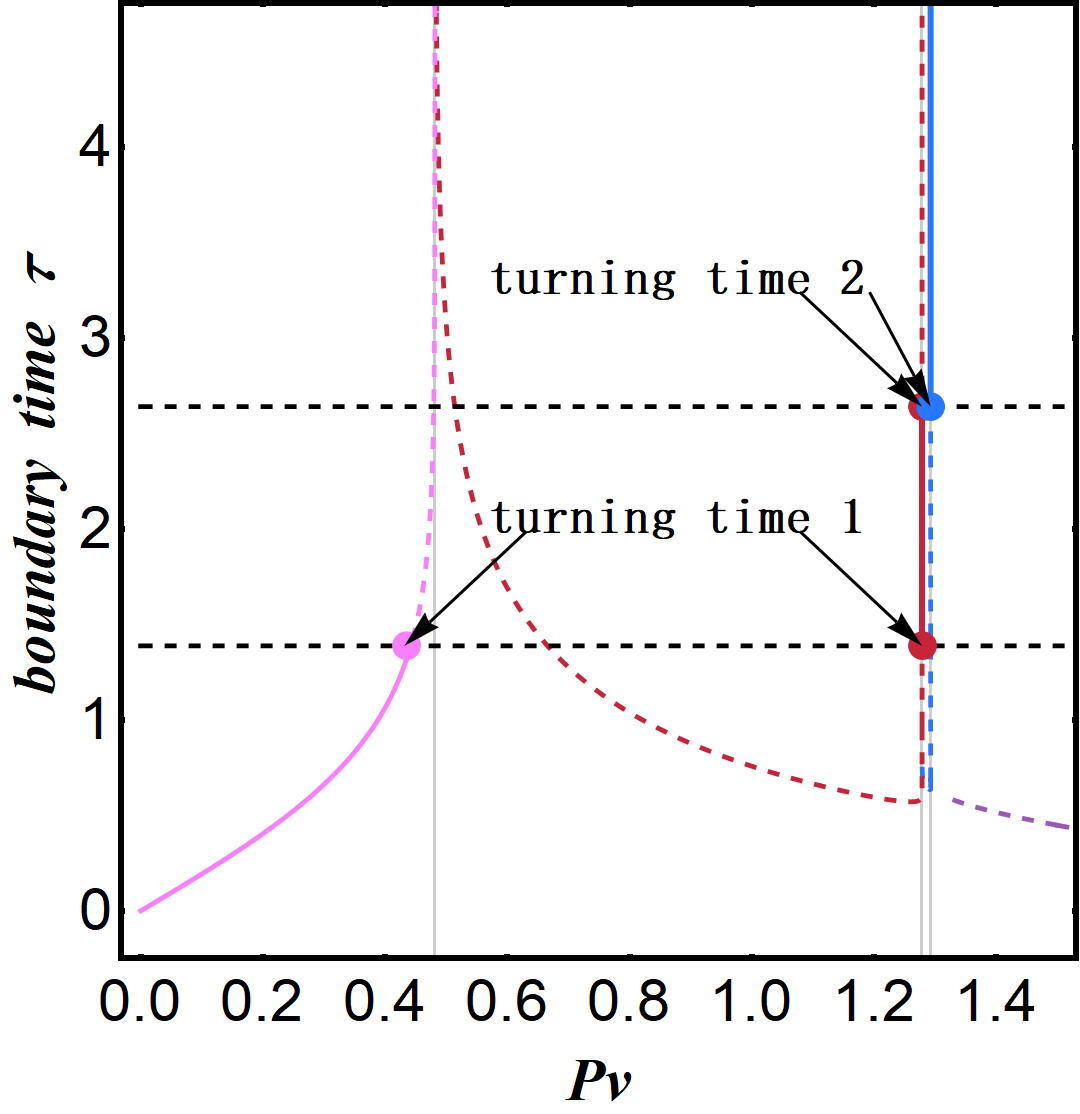}
		\caption{Boundary time $\tau$.}
		\label{SWXUR32b}
	\end{subfigure}
    \caption{{(a): The image of the effective potential $U(r)$ with $L=1$, $\tilde{\lambda}_{1}=10^{-2}$, $\tilde{\lambda}_{2}=3 \times 10^{-6}$ and $\tilde{\lambda}_{3}=1.967 \times 10^{-10}$.
            (b): The relation between the boundary time $\tau$ and the conserved momentum $P_{v}$ using the parameters mentioned above.
            The three solid grey lines represent the three local maxima of $U(r)$, i.e., $U(r_{f})$.
            The pink, red and blue curves correspond to the right, middle and left peaks respectively, and the purple curve corresponds to the divergent branch on the left.
            The pink, red and blue solid lines represent larger generalized volumes at the corresponding boundary time, while the dotted lines are the opposite.}}
	\label{SWXUR32}
\end{figure}

Similarly, we can also consider polynomials with three monomials or even more. Choosing $N=3$, we can get
\begin{align}
    a(r)&=1-\tilde{\lambda}_{1}\left(\frac{r}{L}\right)^{2d}+\tilde{\lambda}_{2}\left(\frac{r}{L}\right)^{4d}-\tilde{\lambda}_{3}\left(\frac{r}{L}\right)^{6d},\\
    U(r)&=-r^{-d}\left(\frac{r}{L}\right)^{2d}(r^{d}-r_{h}^{d})
    \left[1-\left(\frac{r_{h}}{r}\right)^{2d}\tilde{\lambda}_{1}+\left(\frac{r_{h}}{r}\right)^{4d}\tilde{\lambda}_{2}-\left(\frac{r_{h}}{r}\right)^{6d}\tilde{\lambda}_{3}\right]^{2}.
\end{align}
As before, we continue to consider the planer black hole in $(3+1)$-dimensional spacetime.
Choosing different values of the parameters, we may have from $0$ to $3$ local maxima of the effective potential, as shown in the left image of Fig.~\ref{SWXUR3}.
In this case, there are three growth rates at late time.
However, the number of turning times does not necessarily have to be two, as shown in the right image of Fig.~\ref{SWXUR3}.
The highest peak is so steep that the middle peak corresponds to a generalized volume that can never be the largest.
On the other hand, if we select more extreme parameters such that the heights of the two peaks on the left and middle are very close to each other, we can identify a second turning time, as shown in Fig.~\ref{SWXUR32}.

    The emergence of the turning time shows that the growth rate of the complexity is
    abrupt while the complexity itself is still continuous. This phenomenon may imply that
    there are phase transitions, and may means that the shortest geodesic between the reference
    state and the target state will change. 
	{Similar phase transitions have also been observed in quantum computation complexity~\cite{Caputa:2022eye,Suzuki:2023wxw,Roca-Jerat:2023mjs,Anegawa:2024wov} and holographic complexity~\cite{Yang:2023qxx,Auzzi:2023qbm,Wang:2023ipy}.}

\subsection{Five-Dimensional Gauss-Bonnet-AdS Black Hole}\label{secGB}
\begin{figure}[htbp]
	\centering
	\begin{subfigure}[b]{0.45\textwidth}
		\centering
		\includegraphics[scale=0.17]{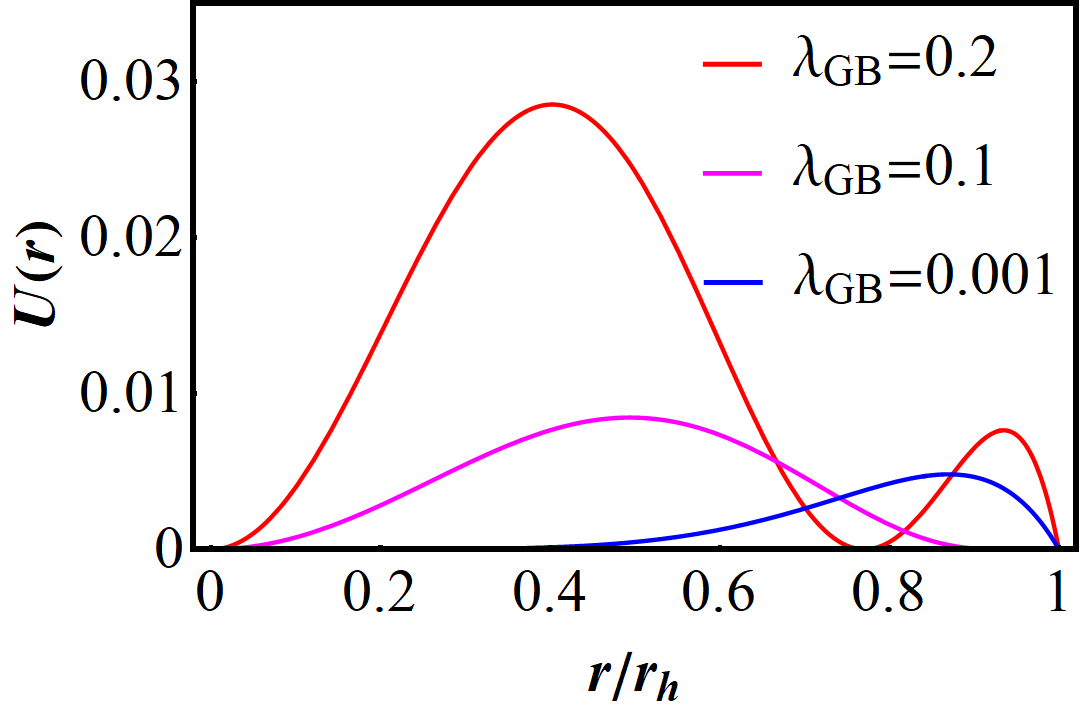}
		\caption{Effective potential $U(r)$.}
		\label{GBUrraa}
	\end{subfigure}
    \hspace{0.05\textwidth}
	\begin{subfigure}[b]{0.45\textwidth}
		\centering
		\includegraphics[scale=0.17]{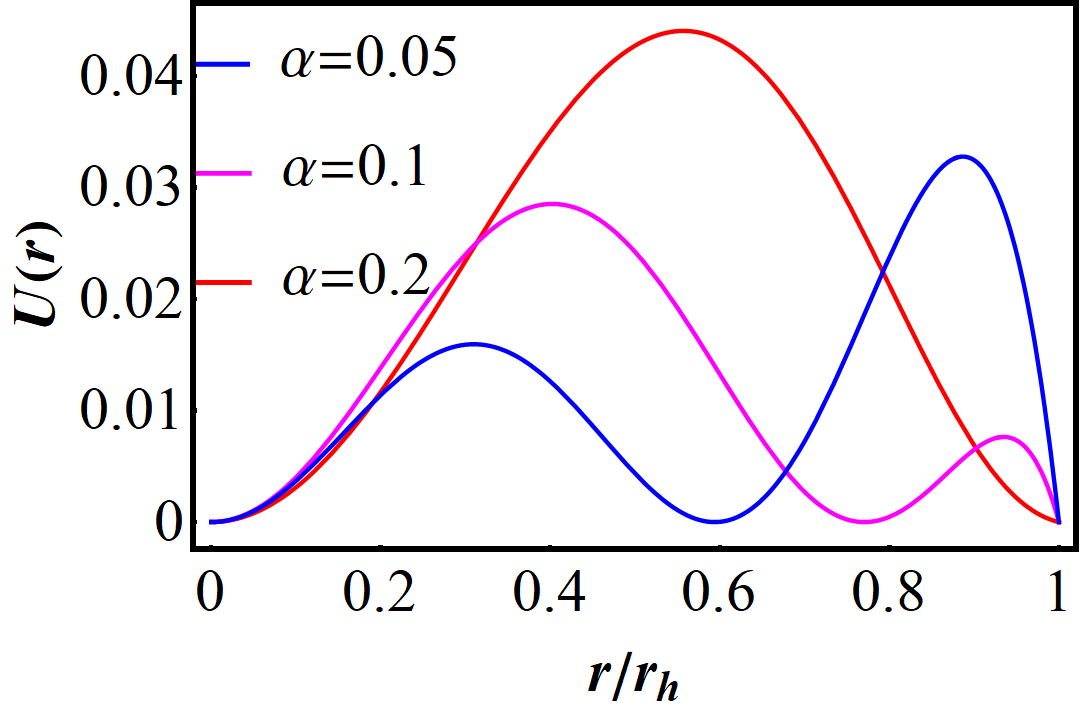}
		\caption{Effective potential $U(r)$.}
		\label{GBUrrab}
	\end{subfigure}
    \caption{The shape of the effective potential. 
	(a): $L=1, m=1, \alpha=0.1$. 
	(b): $L=1, m=1, \lambda_{GB}=0.2$.}
	\label{GBUrra}
\end{figure}

In Ref.~\cite{Wang:2023noo}, the authors analyzed the four-dimensional and five-dimensional Gauss-Bonnet-AdS black hole, 
and found that there is a turning time for the four-dimensional Gauss-Bonnet-AdS black hole, which has two horizons and the extremal hypersurfaces are confined between them. 
In this subsection, we will delve deeper into this issue and examine the distinctions among various gravitational
observables. The metric of the five-dimensional Gauss-Bonnet-AdS black hole is given by Ref.~\cite{Cai:2001dz}
\begin{equation}
	ds^{2}=-f(r)dv^{2}+2dv dr+r^{2}d\Omega_{3}^{2},
\end{equation}
where
\begin{equation}
	f(r)=1+\frac{r^{2}}{2\alpha}\left[1-\sqrt{1+8\alpha\left(\frac{m}{r^{4}}-\frac{1}{L^{4}}\right)}\right],
\end{equation}
$\alpha$ is a coupling constant with dimensions of length squared and $m\equiv 16\pi G M/(d-2)\Omega_{d-2}$ represents the re-scaled mass of the black hole. 
We can easily obtain the effective potential, i.e. 
\begin{equation}
	U(r)=-f(r)a^{2}(r)r^{6}.
	\label{eqgbur}
\end{equation}
In this section, we use the Gauss-Bonnet invariant and the Weyl tensor respectively to
construct $a(r)$. For the Gauss-Bonnet case, we choose
\begin{equation}
	a_{GB}(r)=1+\lambda_{GB} \alpha^{2} \mathcal{R}_{GB},
	\label{agb}
\end{equation}
where $\lambda_{GB}$ is the dimensionless coupling constant, and $\mathcal{R}_{GB}$ is given by 
\begin{equation}
	\mathcal{R}_{GB}=\frac{12(2(f(r)-1))f(r)'+r f(r)'^{2}+r(f(r)-1)f(r)''}{r^{3}}.
\end{equation}
The effective potential \eqref{eqgbur} calculated with $a(r)_{GB}$ is
shown in Fig.~\ref{GBUrra}. 

\begin{figure}[htbp]
	\centering
	\begin{subfigure}[b]{0.45\textwidth}
		\centering
		\includegraphics[scale=0.17]{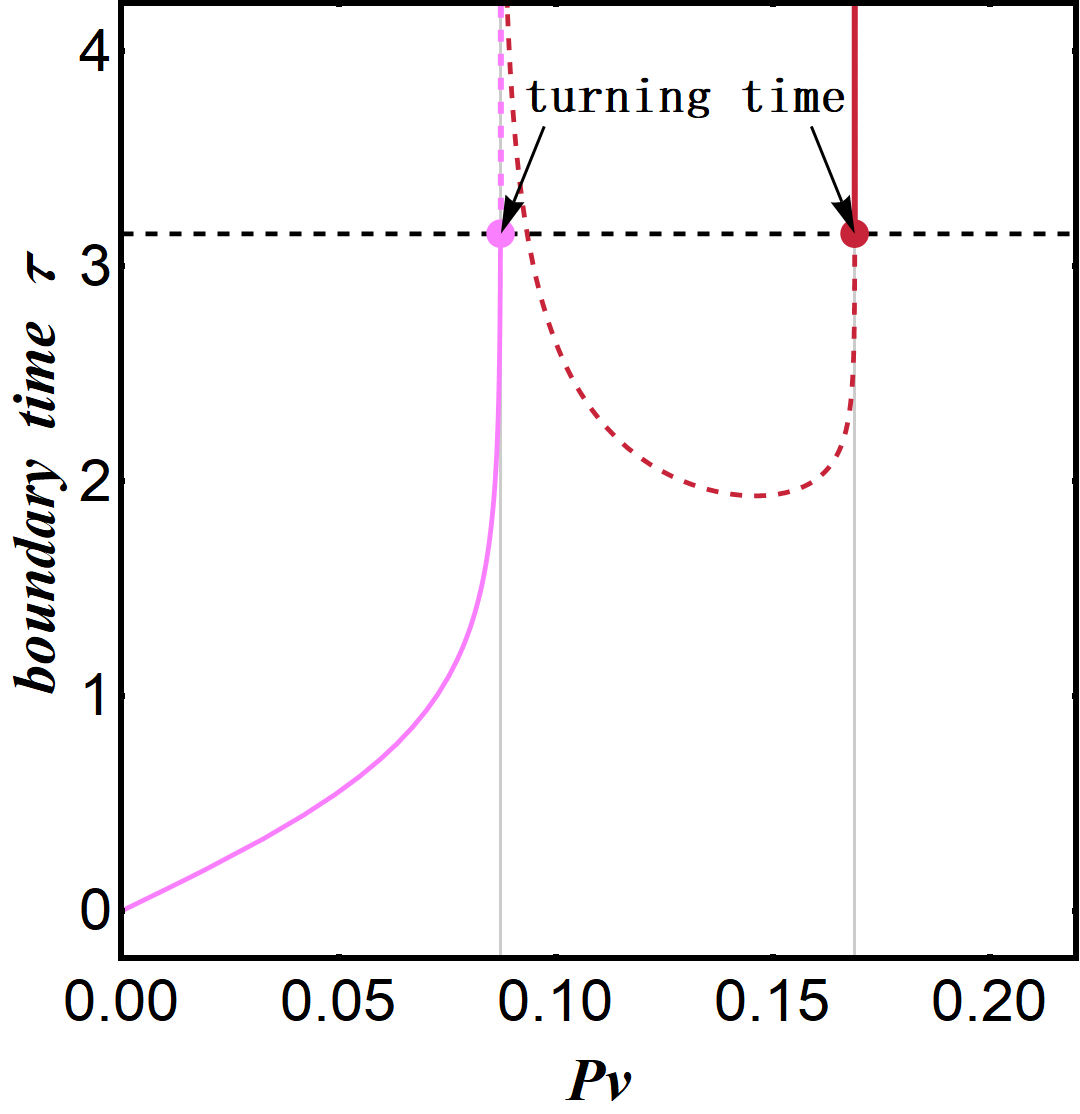}
		\caption{Boundary time $\tau$.}
		\label{GBpvta}
	\end{subfigure}
    \hspace{0.05\textwidth}
	\begin{subfigure}[b]{0.45\textwidth}
		\centering
		\includegraphics[scale=0.17]{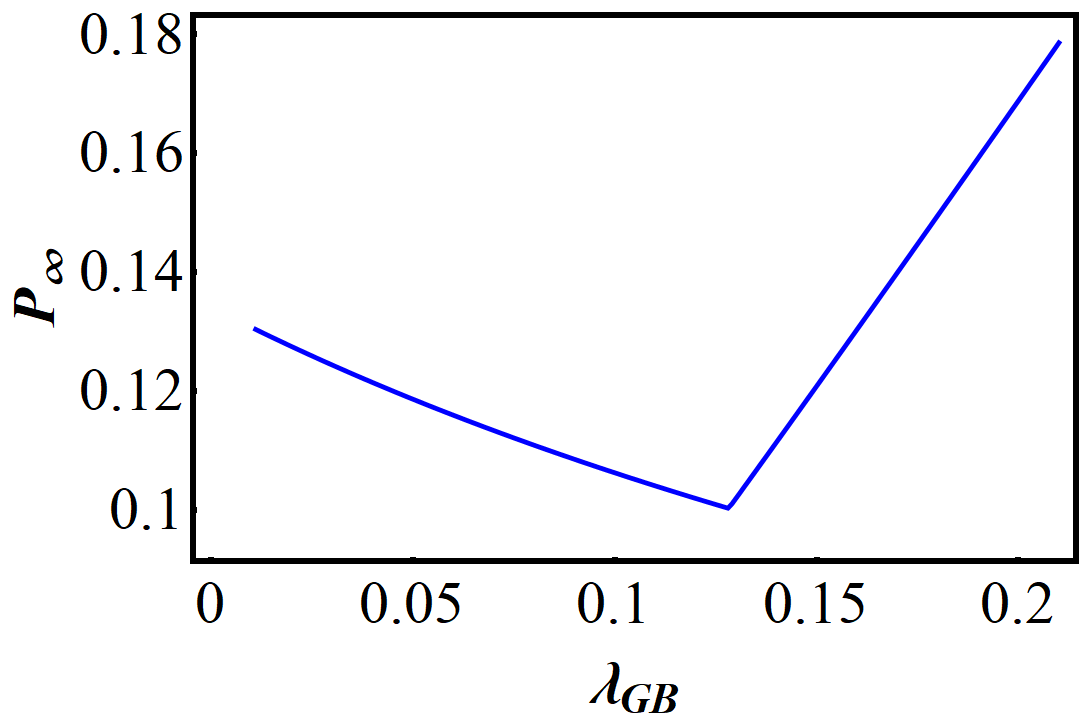}
		\caption{Growth rate at late time. }
		\label{GBpvtb}
	\end{subfigure}
    \caption{(a): The relation between the boundary time $\tau$ and the conserved momentum $P_{v}$ with
	 	$L=1,~m=1,~\alpha=0.1,~\lambda_{GB} = 0.2$. 
		(b): The relation between the coupling constant $\lambda $and the conserved
	 	momentum at late time $P_{\infty}$ with $L=1,~m=1,~\alpha=0.1$.}
	\label{GBpvt}
\end{figure}

In this case, the effective potential goes to zero at the singularity. Additionally, since it remains consistently
zero on the event horizon and the effective potential is always non-negative in the space-like region inside the black hole, there is at least one extreme hypersurface at late time. 
By selecting the appropriate coupling constant $\lambda_{GB}$, we can find the turning time as shown in the left
image of Fig.~\ref{GBpvt}. It is worth noting that in this case, the growth rate of the complexity at
late time always exist and is not zero. It has a lower limit when the turning time tends to
infinity, as shown in the right image of Fig.~\ref{GBpvt}.

\begin{figure}[htbp]
	\centering
	\begin{subfigure}[b]{0.45\textwidth}
		\centering
		\includegraphics[scale=0.22]{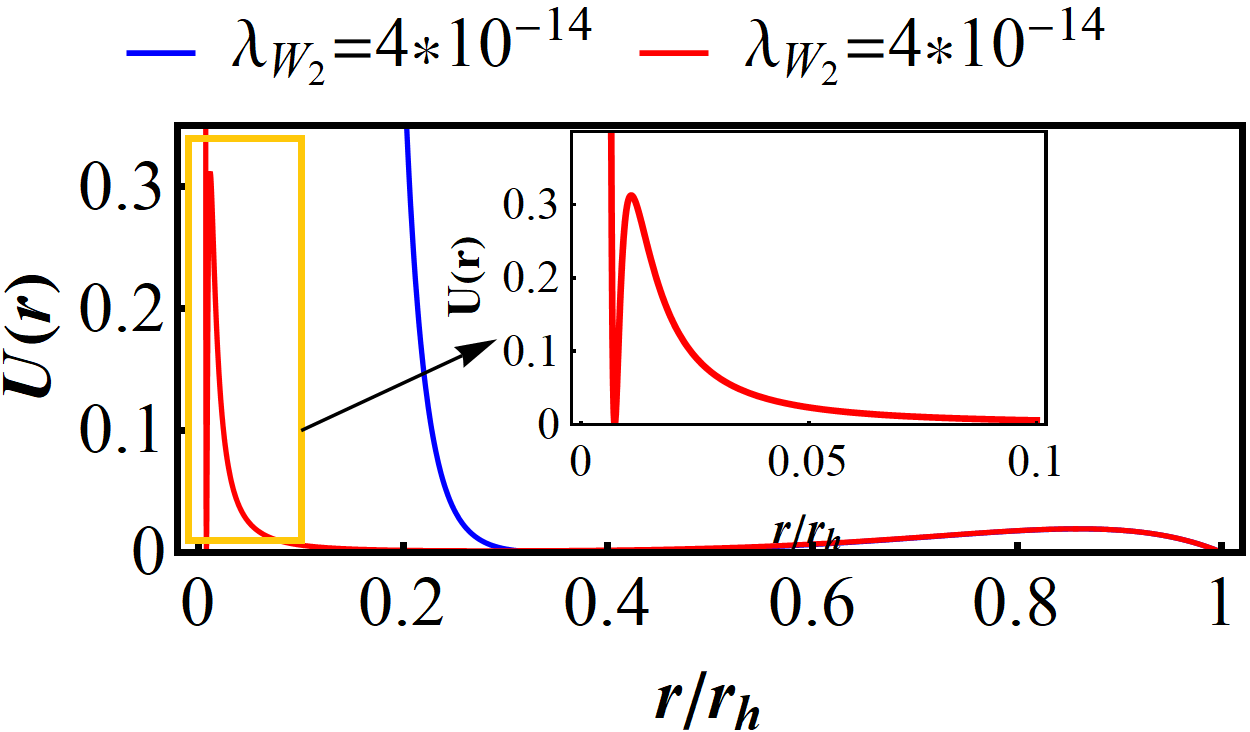}
		\caption{Effective potential $U(r)$.}
		\label{GBC2Ua}
	\end{subfigure}
    \hspace{0.05\textwidth}
	\begin{subfigure}[b]{0.45\textwidth}
		\centering
		\includegraphics[scale=0.17]{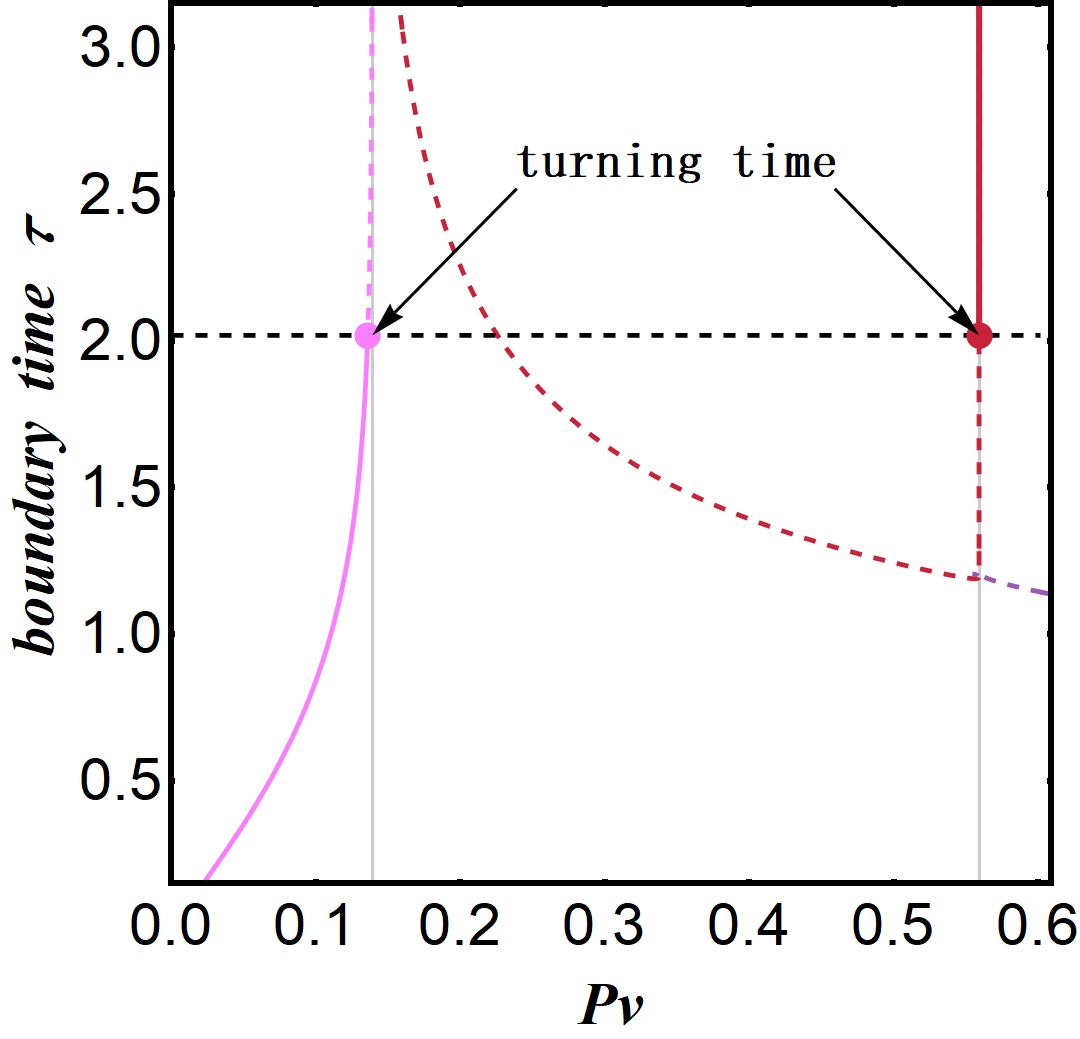}
		\caption{Boundary time $\tau$.}
		\label{GBC2Ub}
	\end{subfigure}
    \caption{(a): The shape of the effective potentia with $L=1,~m=1,~\alpha=0.1,~\lambda_{W_{1}} = 5*10^{-4}$. 
	The red line gives the two peaks that are larger on the left than on the right, the blue line has only one peak, and the black
	dotted line has no peak.
	(b): The relation between the boundary time $\tau$ and the conserved momentum $P_{v}$ with $L=1,~m=1,~\alpha=0.1,~\lambda_{W_{1}} = 5*10^{-4},~\lambda_{W_{1}} = 4*10^{-14}.$}
	\label{GBC2U}
\end{figure}

For the Weyl tensor case, there are some discussions in Ref.~\cite{Wang:2023noo}, and we can still find the turning time by adding higher order terms, i.e.
\begin{equation}
	a_{W}(r)=1+\lambda_{W_{1}}L^{4}C^{2}-\lambda_{W_{2}}L^{8}C^{4},
	\label{aweyl}
\end{equation}
where
\begin{equation}
	C^{2}=\frac{(2-2f(r)+2rf(r)'-r^{2}f(r)'')^{2}}{2r^{4}}.
\end{equation}
In this case, the effective potential diverges at a singularity, as shown in the left image of Fig.~\ref{GBC2U}. We can still find the turning time by choosing an appropriate coupling constant $\lambda_{W}$, as shown in
the right image of Fig.~\ref{GBC2U}.


For the case where the effective potential is zero at a singularity, the presence of
extremal hypersurfaces that cannot reach the late-time regime will no longer occur. As
the effective potential is consistently zero at the horizon, if it is also zero at the singularity,
there must be at least one local maximum between the horizon and the singularity. This
situation is similar to a black hole with two horizons. And when the effective potential
diverges at the singularity, the existence of the local maximum depends on the choice of
the coupling constant.

In our opinion, the generalized volume-complexity can be divided into two categories.
In the first category, the effective potential diverges at the singularity, while in the second one,
it either goes to zero at the singularity or the Cauchy horizon. For the single horizon
black holes without singularity, there exists a third possible scenario for the effective
potential, which we consider to be the same as when it diverges at the singularity, and thus
belong to the first category, see appendix~\ref{appa} for details. The first category will make the
coupling constant has a limit, which requires that there exists an extremal hypersurface at late time. 
In the second category, the coupling constants are desirable across the full phase space, as the extreme hypersurfaces capable of evolving at late time always exist.
Different values of the coupling constant will only affect the growth rate of the complexity
at late time, as well as the presence or absence of the turning time. In this case, the growth
rate of the generalized volume-complexity will have a lower limit value, but not zero.

\section{Holographic Complexity for Black Holes with Two Horizons}\label{sec4}
In this section, we take the charged BTZ black hole as an example to calculate its generalized volume-complexity for black holes with a Cauchy horizon.
In this case, there is no generalized volume-complexity of the first category. 
We will respectively use space-time curvature and matter to construct the gravitational observables.
Let us first give the charged BTZ black hole, whose metric is given by Refs.~\cite{Banados:1992wn,Martinez:1999qi},
\begin{align}
    ds^{2}&=-f(r)dt^{2}+\frac{dr^{2}}{f(r)}+r^{2}d\theta^{2},\notag\\
    f(r)&=-M+\frac{r^{2}}{L^{2}}-\pi Q^{2}\ln{\frac{r}{L}}
\end{align}
The metric is asymptotically AdS$_{3}$ with radius $L$. By convention, we set $8G_{N}\equiv 1$, where $G_{N}$ is the Newton's constant.
The mass and the charge of the black hole are denoted by $M$ and $Q$ respectively, and can be expressed in terms of the event horizon and the Cauchy horizon $r_{\pm }$:
\begin{align}
M=&\frac{r_{+}^{2}\ln{\frac{r_{-}}{L}}-r_{-}^{2}\ln{\frac{r_{+}}{L}}}{L^{2}\ln{\frac{r_{-}}{r_{+}}}},\\
Q^{2}=&\frac{r_{+}^{2}-r_{-}^{2}}{\pi L^{2}\ln{\frac{r_{+}}{r_{-}}}}.
\end{align}
We can also rewrite the metric into the ingoing Eddington-Finkelstein coordinates, that is
\begin{equation}
    ds^2=-f(r)dv^2+2dvdr+r^2d\theta^2.
\end{equation}
The generalized volume-complexity can be represented by
\begin{equation}
    \mathcal{C}=2\pi\int d\sigma ~r~\sqrt{2\dot{v}\dot{r}-f(r)\dot{v}^{2}}~a(r)\equiv \frac{2\pi}{G_{N}L}\int_{\Sigma}d\sigma \mathcal{L}(r,\dot{v},\dot{r}).
    \label{CLBTZ}
\end{equation}
Obviously, the gauge given by Eq.~\eqref{gauge} becomes
\begin{equation}
    \sqrt{-f(r)\dot{v}^{2}+2\dot{v}\dot{r}}= a(r) {r}.
\end{equation}
We can just write down its effective potential, complexity, and complexity growth rate:
\begin{align}
    U(r)=&-f(r)a^{2}(r)r^{2},\\
\mathcal{C}=&-\frac{4\pi}{G_{N}L}\int_{r_{min}}^{\infty }\frac{U(r)}{f(r)\sqrt{U(r_{min})-U(r)}}dr,\\
\frac{d\mathcal{C}}{d\tau}=&\frac{2\pi}{G_{N}L}P_{v}.
\end{align}

\subsection{Generalized Volume-Complexity Constructed by Space-Time Curvature}
\begin{figure}[htbp]
	\centering
	\begin{subfigure}[b]{0.45\textwidth}
		\centering
		\includegraphics[scale=0.2]{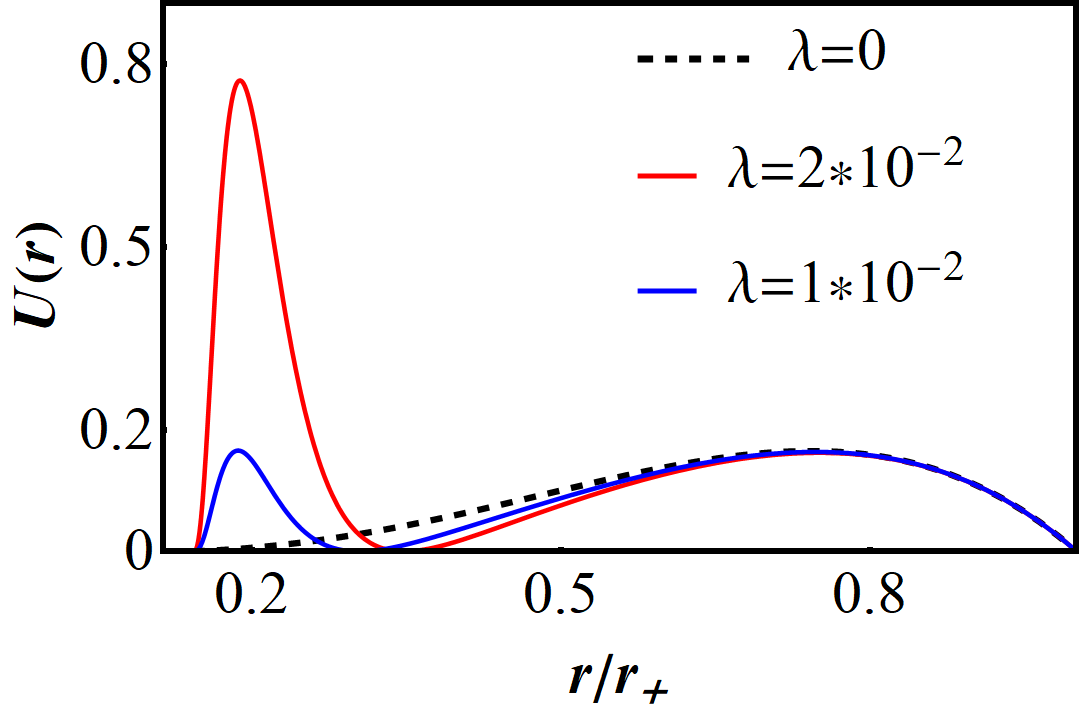}
		\caption{Effective potential $U(r)$.}
		\label{CHBTZC3a}
	\end{subfigure}
    \hspace{0.05\textwidth}
	\begin{subfigure}[b]{0.45\textwidth}
		\centering
		\includegraphics[scale=0.3]{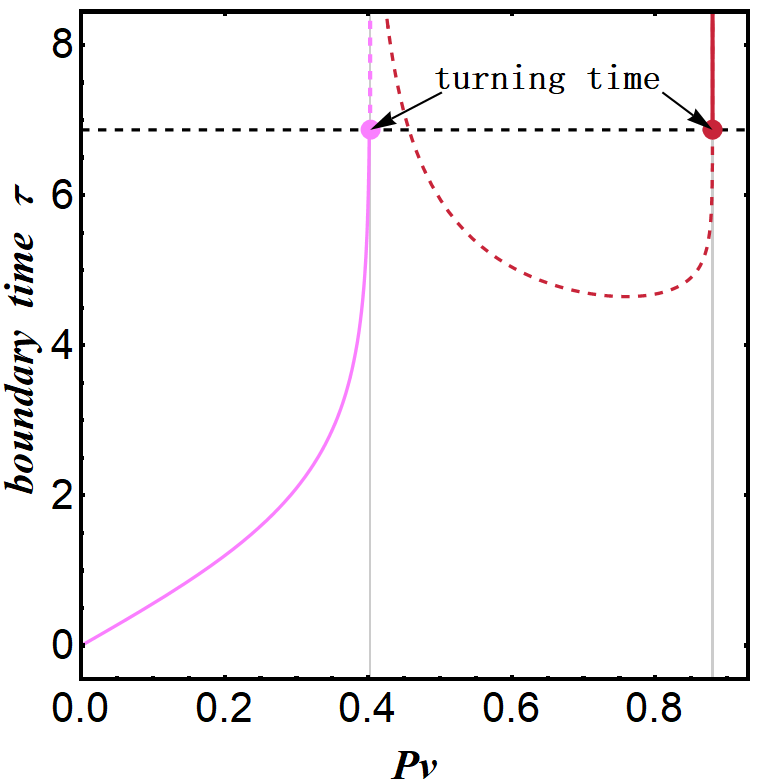}
		\caption{Boundary time $\tau$.}
		\label{CHBTZC3b}
	\end{subfigure}
    \caption{Left: The shape of the effective potential with $L=M=1$ and $Q=0.4$. 
	Right: The relation between the boundary time $\tau$ and the conserved momentum $P_{v}$ 
	with $\lambda=2*10^{-2}$, $L=M=1$, and $Q=0.4$.}
	\label{CHBTZC3}
\end{figure}

It is well known that the Weyl tensor is always zero when $d=2$, so we need to look for other non-trivial scalar functions.
In this regard, we can naturally think of the three-dimensional replacement of the Weyl tensor, i.e. the Cotton tensor $C_{3}$.
In this subsection, we choose $a(r)=1+\lambda L^{6}C_{3}^{2}$, where 
\begin{equation}
C_{3}=C_{ijk}\equiv\nabla_{i}R_{jk}-\nabla_{j}R_{ik}-\frac{1}{4}(\nabla_{i}Rg_{jk}-\nabla_{j}Rg_{ik})
\end{equation}
is the Cotton tensor. 
In the three-dimensional case, $C_{3}^{2}=\frac{1}{4}f(r)f'''(r)^{2}$. 
Thus we can obtain the expression of the effective potential, 
\begin{equation}
	U(r)=\left(M - \frac{r^2}{L^2} + \pi Q^2 \ln{\frac{r}{L}}\right) \left(r + \lambda \frac{L^6 \pi^2 Q^4 (-M + r^2/L^2 - \pi Q^2 \ln{\frac{r}{L}})}{r^5}\right)^2
\end{equation}
Similarly, we can still get a figure of the effective potential and the corresponding turning time as shown in Fig.~\ref{CHBTZC3}. 

When $q=0$, the cotton tensor is zero, generalized volume-complexity regresses to the CV proposal.

\subsection{Generalized Volume-Complexity Constructed by Matter Content}
For charged black holes, another naturally occurring scalar function can effectively construct the generalized volume-complexity. i.e. the electromagnetic field action. 
In this situation, we choose $a(r)=1-\lambda L^{2} F_{\mu\nu}F^{\mu\nu} $, where $F_{\mu\nu}F^{\mu\nu}= -2\frac{Q^{2}}{r^{2}}$, and can obtain an expression for the effective potential, that is 
\begin{equation}
U(r)=\left(r-\frac{2 L^{2} Q^{2} \lambda}{r}\right)^{2}\left(M-\frac{r^{2}}{L^{2}}+\pi Q^{2} \ln{\frac{r}{L}}\right). 
\label{Urchbtz}
\end{equation}
Of course, we can use this configuration to analyze all other charged black holes.

\begin{figure}[htbp]
    \centering
    \includegraphics[scale=0.15]{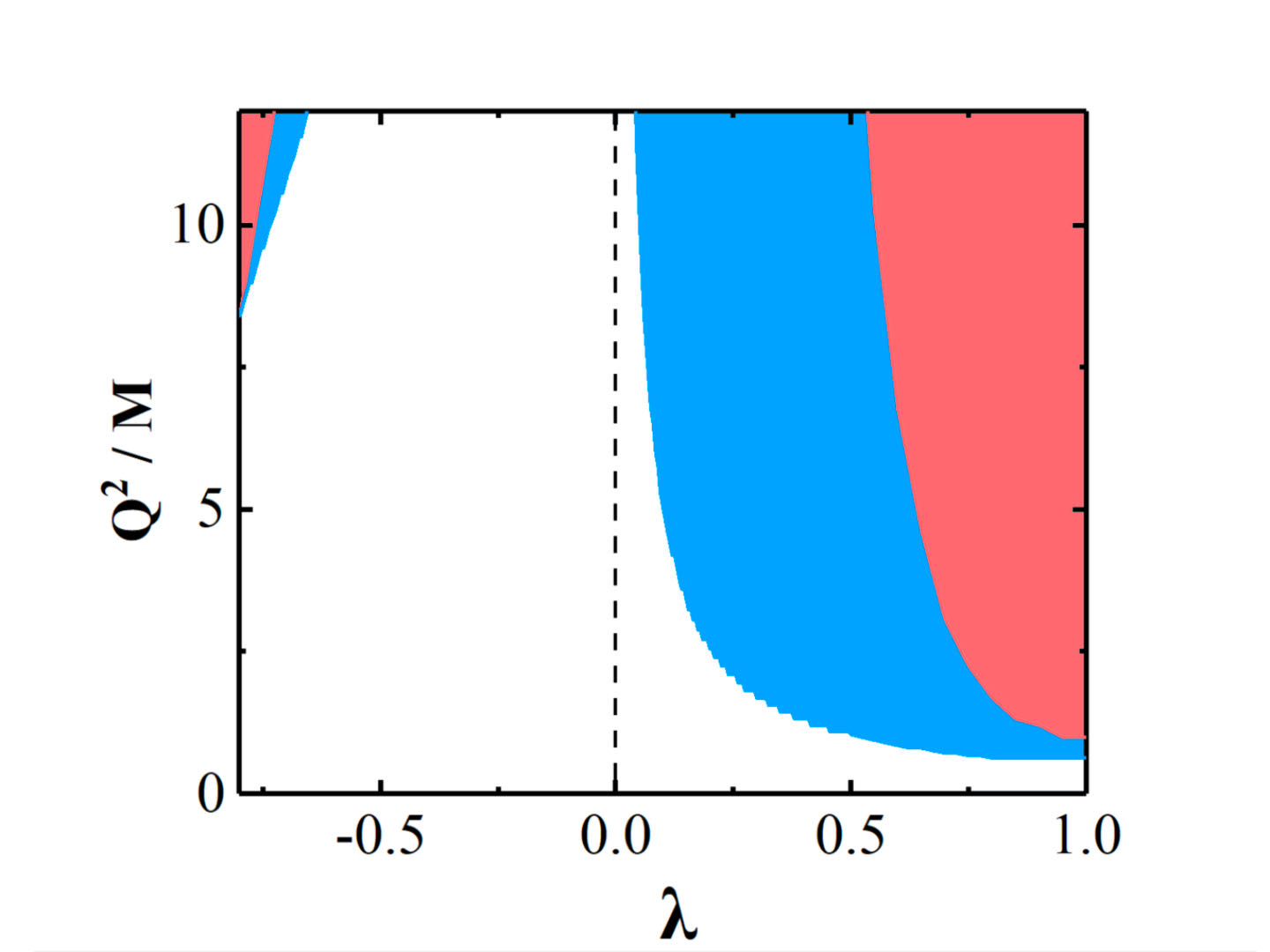}
    \caption[paspbtz]{The parameter space of the effective potential $U(r)$ between the two horizons of the charged BTZ black hole.
    In the red area, there are two local maxima that decrease from left to right. In the blue area, there are also two local maxima, but the local maximum on the right is greater than the one on the left. In the remaining white area, there is only one local maximum. 
   }
    \label{paspbtz}
\end{figure}

\begin{figure}[htbp]
	\centering
	\begin{subfigure}[b]{0.45\textwidth}
		\centering
		\includegraphics[scale=0.18]{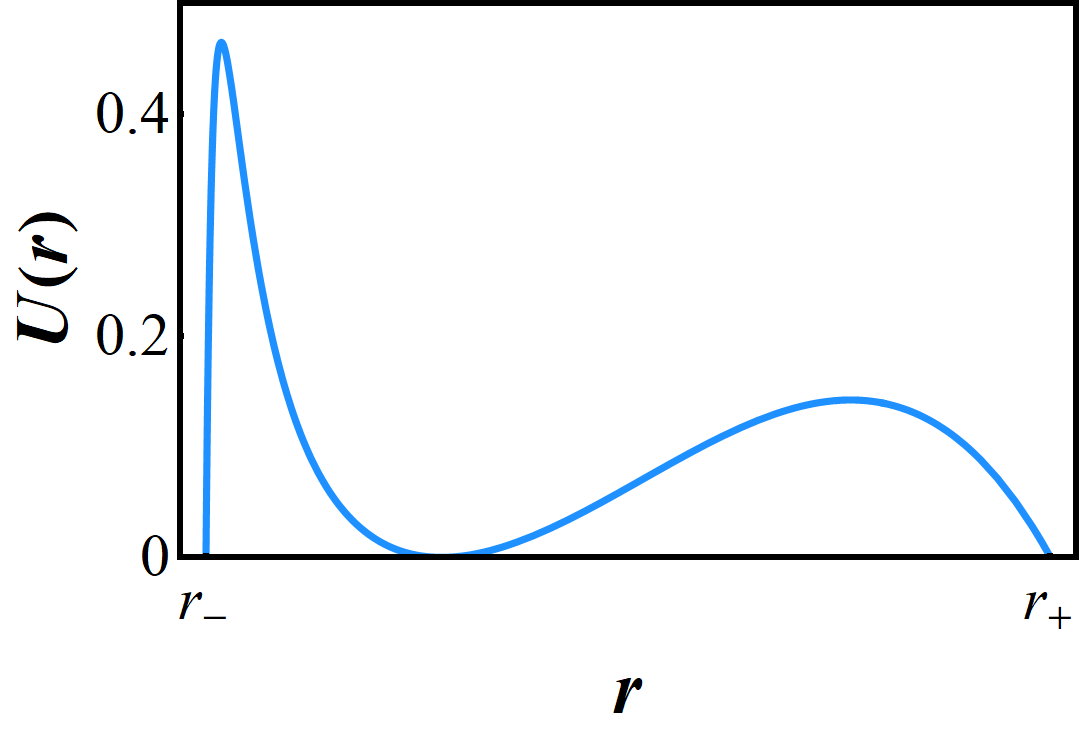}
		\caption{Boundary tims $\tau$.}
		\label{chbtzura}
	\end{subfigure}
    \hspace{0.05\textwidth}
	\begin{subfigure}[b]{0.45\textwidth}
		\centering
		\includegraphics[scale=0.18]{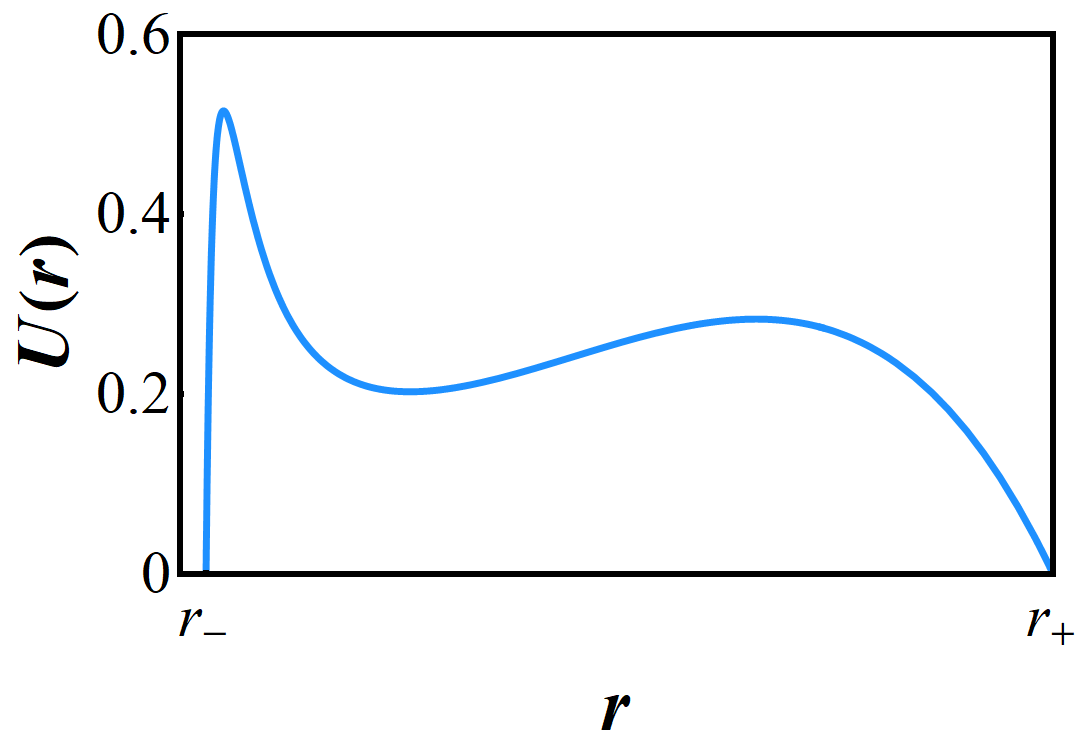}
		\caption{Boundary time $\tau$.}
		\label{chbtzurb}
	\end{subfigure}
    \caption{The effective potential defined in Eq.\eqref{Urchbtz}.
    In this situation, we may have 1 or 2 local maxima of the effective potential for different parameters.
    When the number of the local maxima is one, the result is similar to the CV proposal.
    Therefore, we only consider the case of two local maxima.
    Left: the shape of the effective potential for $\lambda=0.5$, $L=M=1$, and $Q=0.3$. There are two local maxima and the local minimum between them is zero.
    Right: the shape of the effective potential for $\lambda=-0.5$, $L=M=1$, and $Q=0.3$. There are also two local maxima but the minimum between them does not reach zero.
    Whether the local minimum is zero depends only on the positive or negative of $\lambda$. }
	\label{chbtzur}
\end{figure}

\begin{figure}[htbp]
	\centering
	\begin{subfigure}[b]{0.45\textwidth}
		\centering
		\includegraphics[scale=0.23]{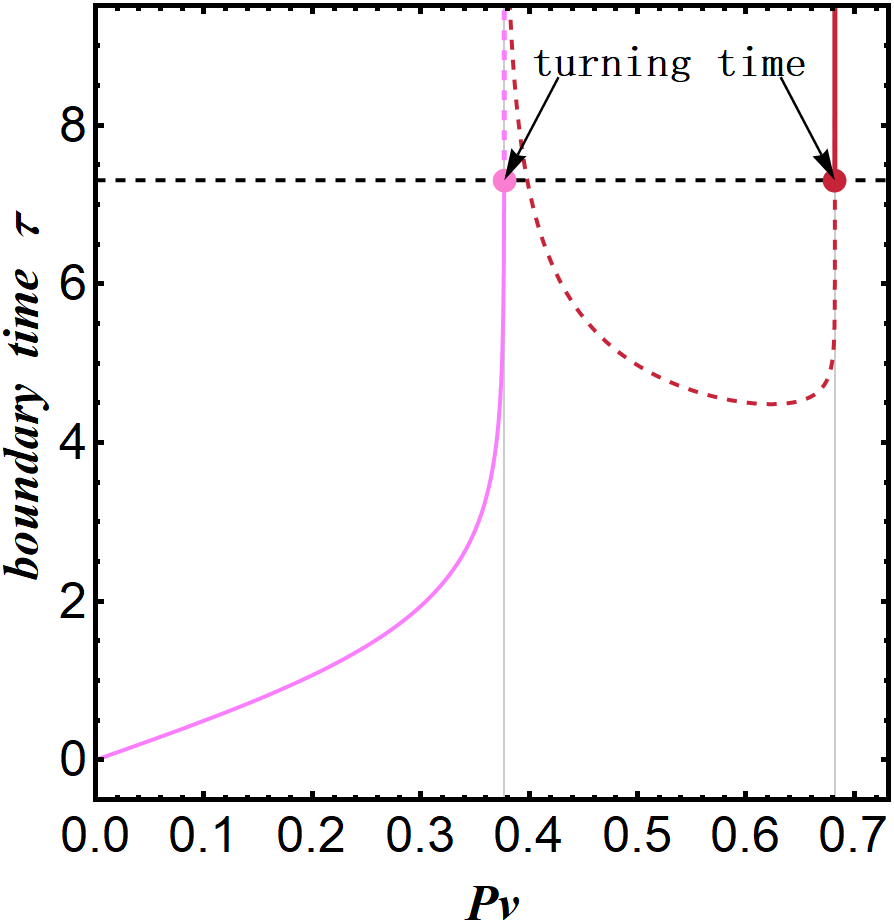}
		\caption{Boundary tims $\tau$.}
		\label{btzturninga}
	\end{subfigure}
    \hspace{0.05\textwidth}
	\begin{subfigure}[b]{0.45\textwidth}
		\centering
		\includegraphics[scale=0.235]{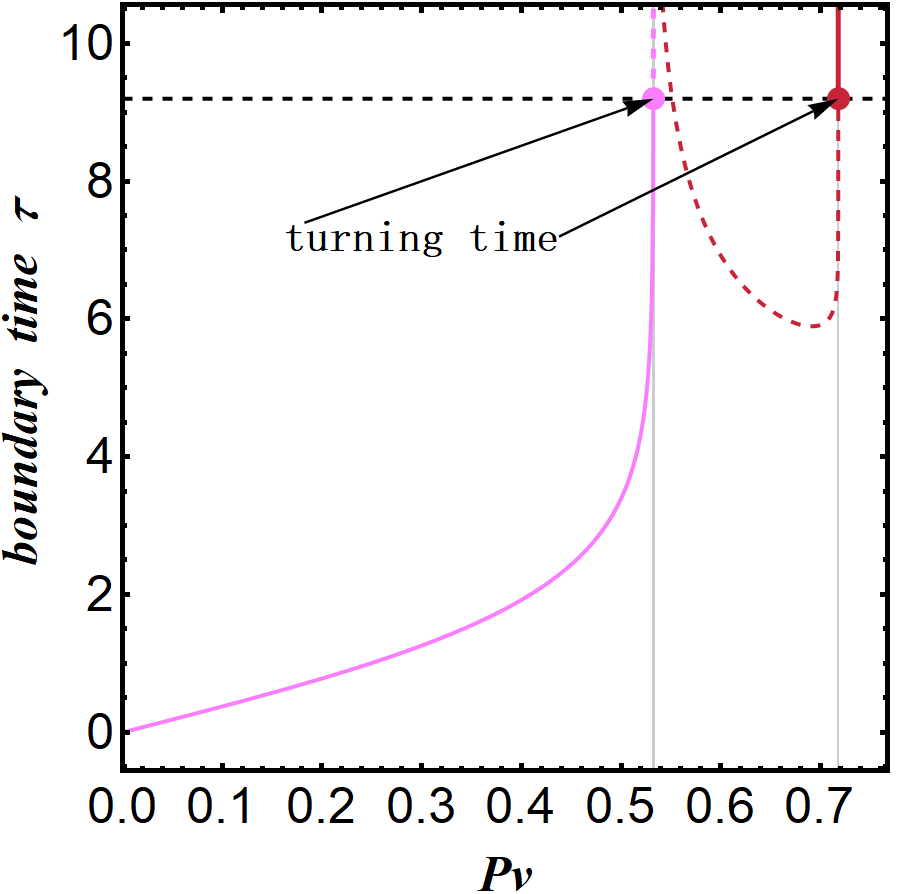}
		\caption{Boundary time $\tau$.}
		\label{btzturningb}
	\end{subfigure}
    \caption[btzturning]{The relation between the boundary time $\tau$ and the conserved momentum $P_{v}$.
    (a): $\lambda=0.5$, $L=M=1$, and $Q=0.3$.
    (b): $\lambda=-0.5$, $L=M=1$, and $Q=0.3$.}
    \label{btzturning}
\end{figure}

In this case, we can still find a series of $\lambda$, $Q$ and $M$ such that $U(r)$
has more than one local maxima between $r_{-}$ and $r_{+}$, the one with smaller $r$ is also larger.
The parameter space is shown in Fig.~\ref{paspbtz}.
In the analysis that follows, we only consider the parameters that fall within the red areas, where we have two local maxima that describe from left to right.
As before, we consider their turning time.
For the example shown in Fig.~\ref{chbtzur}, the boundary time and turning time can be reflected in Fig.~\ref{btzturning}.

\begin{figure}[htbp]
	\centering
	\begin{subfigure}[b]{0.45\textwidth}
		\centering
		\includegraphics[scale=0.17]{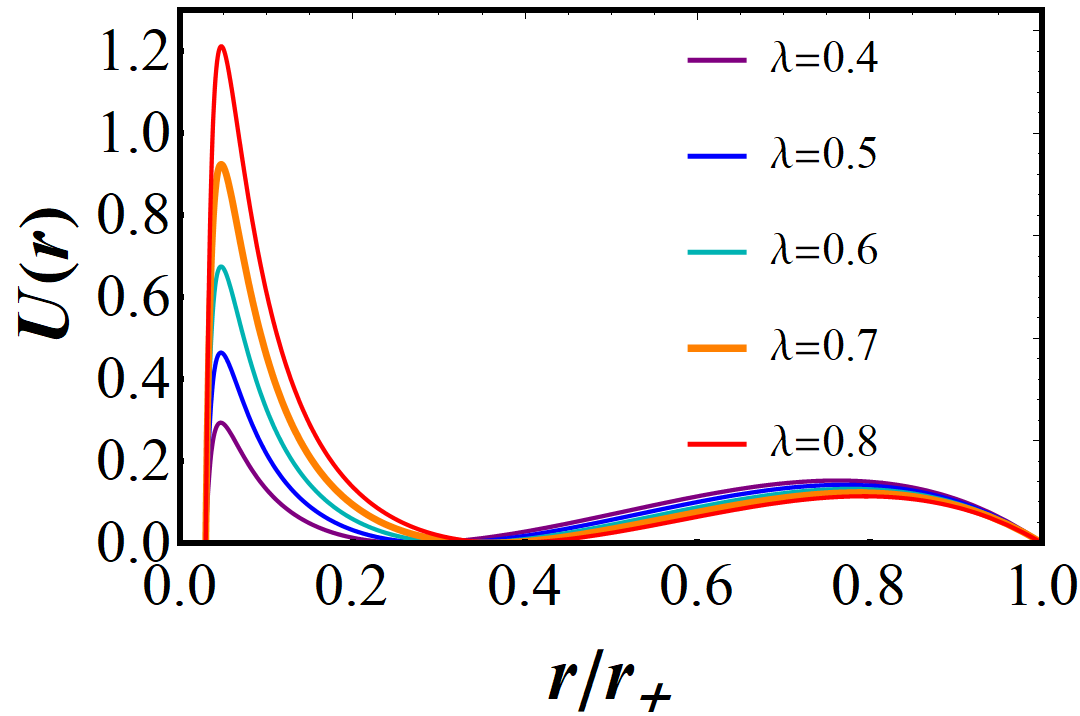}
		\caption{Effective potential $U(r)$.}
		\label{btzurmanya}
	\end{subfigure}
    \hspace{0.05\textwidth}
	\begin{subfigure}[b]{0.45\textwidth}
		\centering
		\includegraphics[scale=0.17]{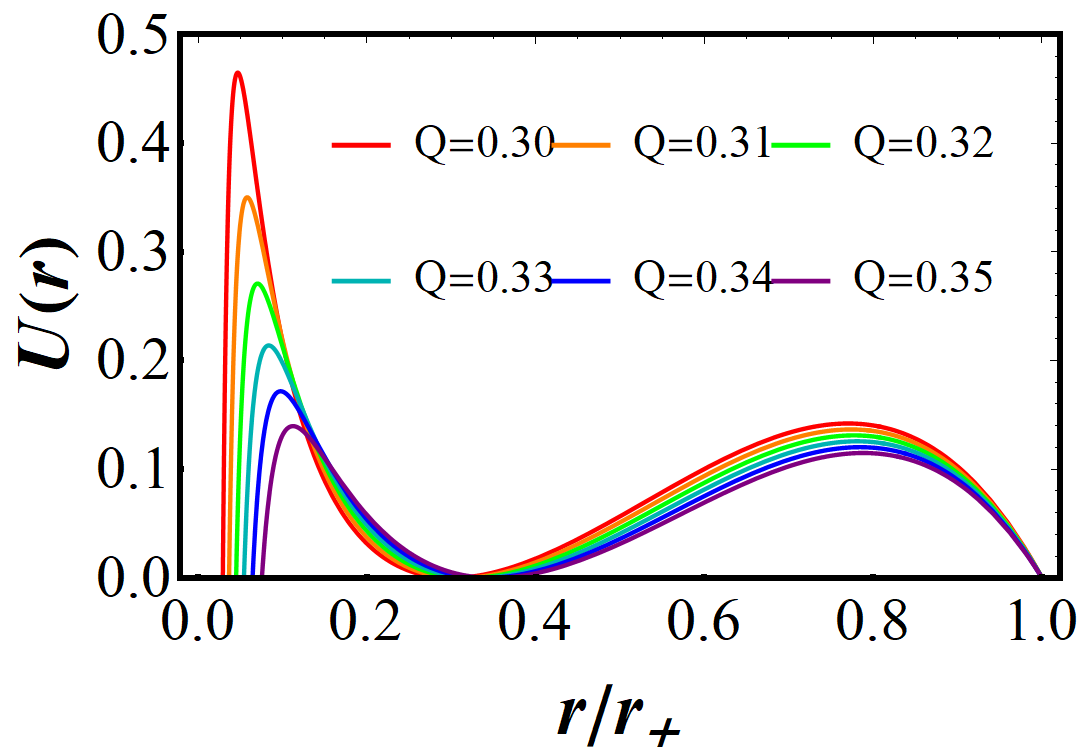}
		\caption{Effective potential $U(r)$.}
		\label{btzurmanyb}
	\end{subfigure}
    \caption[swx]{The shape of the effective potential with  $L=M=1$.
	(a): $\lambda=0.4$ to $0.8$ and $Q=0.3$.
    (b): $Q=0.30$ to $0.35$ and $\lambda=0.5$.}
    \label{btzurmany}
\end{figure}

We fix $M=L=1$ and choose different $\frac{Q^{2}}{M}$ or $\lambda$ to construct a series of $U(r)$ as shown in Fig.~\ref{btzurmany}, where we only consider the case that the effective potential has two loacl maxima.
It is worth mentioning that when we fix $M=L$, $r_{-}$ varies monotonically with $\frac{Q^{2}}{M}$ 
as shown in the right figure of Fig.~\ref{btzurmany}.
The turning time of the complexity is still logarithmic with its parameter. 
\begin{align}
    \tau_{turning}=&A_{QM}-B_{QM}\ln\left(\frac{M}{Q^{2}}+C_{QM}\right),\notag\\
    \tau_{turning}=&A_{\lambda}-B_{\lambda}\ln(\lambda+C_{\lambda}),
\end{align}
where $A$, $B$ and $C$ are the fitting values.
Among them, $A$ and $B$ have length dimensions, and $C$ is a dimensionless constant.
This conclusion is similar to that of constructing generalized volume-complexity with space-time curvature discussed in Sec.~\ref{sec3}.

\subsection{RN-AdS Black Hole Rediscussion} 
\begin{figure}[htbp]
	\centering
	\begin{subfigure}[b]{0.45\textwidth}
		\centering
		\includegraphics[scale=0.28]{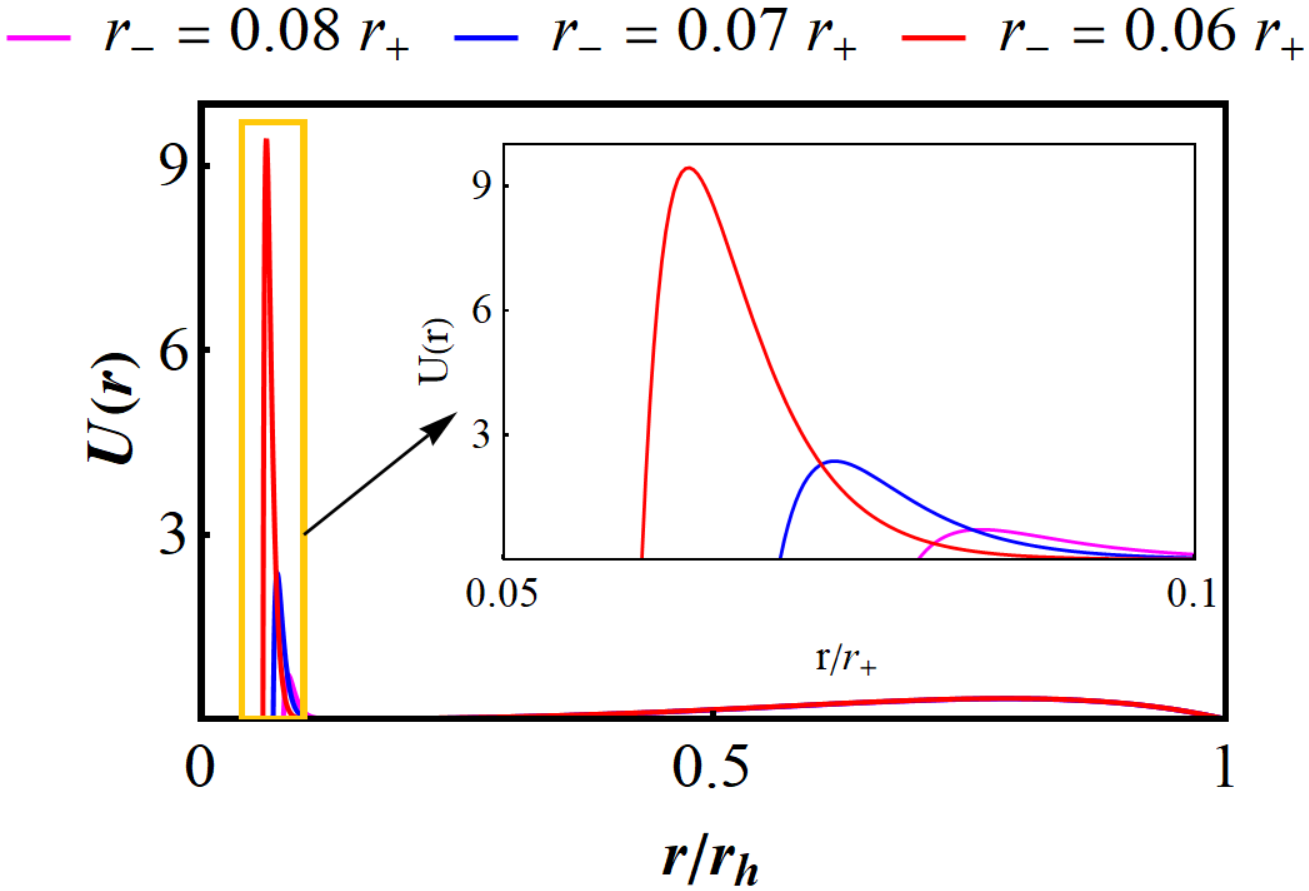}
		\caption{Boundary tims $\tau$.}
		\label{RNUra}
	\end{subfigure}
    \hspace{0.05\textwidth}
	\begin{subfigure}[b]{0.45\textwidth}
		\centering
		\includegraphics[scale=0.19]{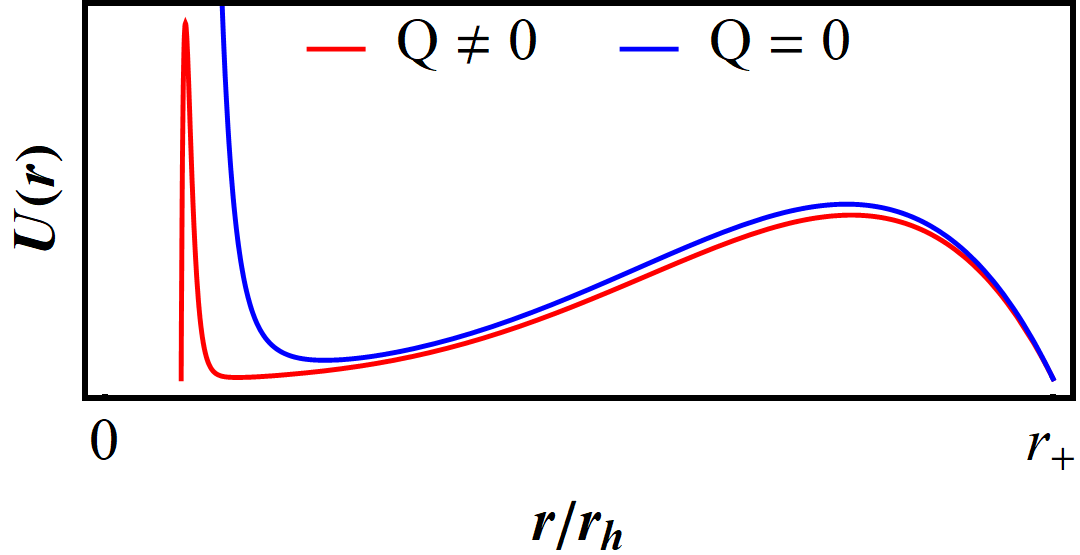}
		\caption{Boundary time $\tau$.}
		\label{RNUrb}
	\end{subfigure}
	\caption[123]{(a): The shape of the effective potential with $L=r_{+}=1$, $\lambda=1*10^{-6}$.
	(b): The comparison of $Q \to 0$ and $Q = 0$.}
	\label{RNUr}
\end{figure}

In our previous work~\cite{Wang:2023eep}, we considered the four-dimensional RN-AdS black hole as an example to find the turning time. 
However, it is important to note that the conclusion regarding the RN-AdS black hole does not necessarily converge to that of the AdS-Schwarzschild black hole. 
Therefore, we continue our research using the metric under the Eddington-Finkelstein coordinate, denoted as 
\begin{align}
	ds^2=&-f(r)dv^2+2dvdr+r^2d\Omega ^{2},\\
	f(r)=&1-\frac{2M}{r}+\frac{Q^2}{r^2}+\frac{r^2}{L^2} \notag \\
  =&\frac{(r-r_1)(r-r_2)(L^2 +r^2 +r_{1}^{2} +r_{2}^{2} +r_1 r_2 +r(r_1 +r_2))}{r^2 L^2},
\end{align}
where
\begin{align}
	M=&\frac{(r_1+r_2)(L^2+r_{1}^{2}+r_{2}^{2})}{2L^2},\\
	Q=&\frac{\sqrt{ r_1 r_2 (L^2+r_{1}^{2}+r_1 r_2+r_{2}^{2})}}{L}.
\end{align}
The solution to the equation of motion remains unchanged from before, and we can get the expression of the effective potential $U(r)=-f(r)a^{2}(r)r^{4}$. 
We still use the Weyl tensor to construct the gravitational observable, i.e. 
\begin{align}
	a(r)=&1+\lambda L^4C^2,\\
	C^{2}=&\frac{(2-2f(r)+2rf'(r)-r^{2}f''(r))^{2}}{3 r^{4}}.
\end{align}
Now, we can fix the event horizon and adjust $Q$ so that the Cauchy horizon gradually approaches the singularity, meanwhile the peak on the left also approaches the singularity, as shown in the left image of Fig.~\ref{RNUr}.
When $Q \to 0$, the Cauchy horizon will be infinitely close to the singularity.
At the same time, the left peak still has a finite maximum value that depends on the coupling constant $\lambda$.
However, if $Q = 0$, the black hole returns to the Schwarzschild-AdS black hole. 
With the disappearance of the Cauchy horizon, the left peak disappears together and the effective potential diverges at the singularity. 
In other words, in the configuration we are considering, as the charge $Q$ disappears, the second category of the generalized volume-complexity transitions to the first case. 
This mutation in the growth rate of complexity at late time is shown in the right image of Fig.~\ref{RNUr}.
This brings us back to the Schwarzschild situation, where the coupling constant no longer holds in the full parameter space.
{\section{$F_{1} \neq F_{2}$} \label{sec5}
In this section, we will consider the case where $F_{1}\neq F_{2}$. 
We determine the extreme hypersurface by setting $F_{2}=a_{2}(r)$, 
and consider the gravitational observables constructed using $F_{1}=a_{1}(r)$, as shown in Eqs.~\eqref{ef1} and \eqref{ef2}.
Taking the $(d+1)$-dimensional AdS spherically symmetric black hole as an example, we can still use the method in Sec.~\ref{sec2} to get the conserved momentum and effective potential, i.e.
\begin{equation}
	\dot{r}^{2}+U_{2}(r)=P_{v}^{2},~~~~U_{2}(r)=-f(r)a_{2}^{2}(r)r^{2(d-1)}.
\end{equation}
The anchored boundary time $\tau$ is uniquely determined by the properties of the hypersurface, so $\tau$ is only related to $a_{2}(r)$, i.e.,
\begin{align}
	\tau=&-2\int_{r_{min}}^{\infty}dr\frac{P_{v}}{f(r)\sqrt{P_{v}^{2}-U_{2}(r)}}\notag\\
	    =&-2\int_{r_{min}}^{\infty}dr\frac{\sqrt{U_{2}(r_{min})}}{f(r)\sqrt{U_{2}(r_{min})-U_{2}(r)}}.
    \label{tau12}
\end{align}
However, the generalized volume-complexity will depend on both $a_1(r)$ and $a_2(r)$, that is
\begin{equation}
	\mathcal{C}_{F_{1},\Sigma_{F_{2}}}=\frac{2V_{d-1}}{G_{N}L}\int_{r_{min}}^{\infty}\frac{a_{1}(r)a_{2}(r)r^{2(d-1)}}{\sqrt{P_{v}^{2}+f(r)a_{2}^{2}(r)r^{2(d-1)}}}dr.
	\label{C12}
\end{equation}
\subsection{The Restriction of $a(r)$}
We still take the five-dimensional Gauss-Bonnet-AdS black hole as an example, 
and select Eq.~\eqref{agb} and Eq.~\eqref{aweyl} as $a_{1}(r)$ and $a_{2}(r)$, respectively.
Using Eq.~\eqref{tau12} and Eq.~\eqref{C12}, we find that the generalized volume-complexity loses the property of late linear growth, as shown in Fig.~\ref{CtauW}, we think this is unreasonable.
\begin{figure}[htbp]
	\centering
	\begin{subfigure}[b]{0.45\textwidth}
		\centering
		\includegraphics[scale=0.16]{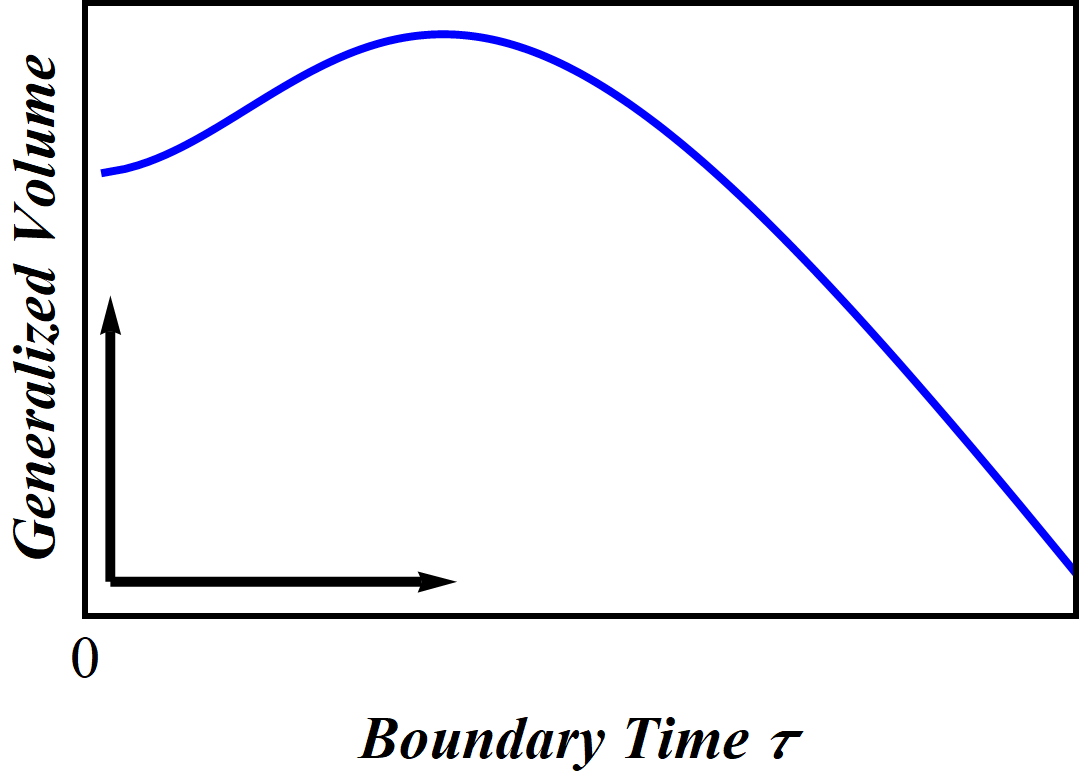}
		\caption{The surface far from the singularity.}
		\label{CtauWa}
	\end{subfigure}
    \hspace{0.05\textwidth}
	\begin{subfigure}[b]{0.45\textwidth}
		\centering
		\includegraphics[scale=0.16]{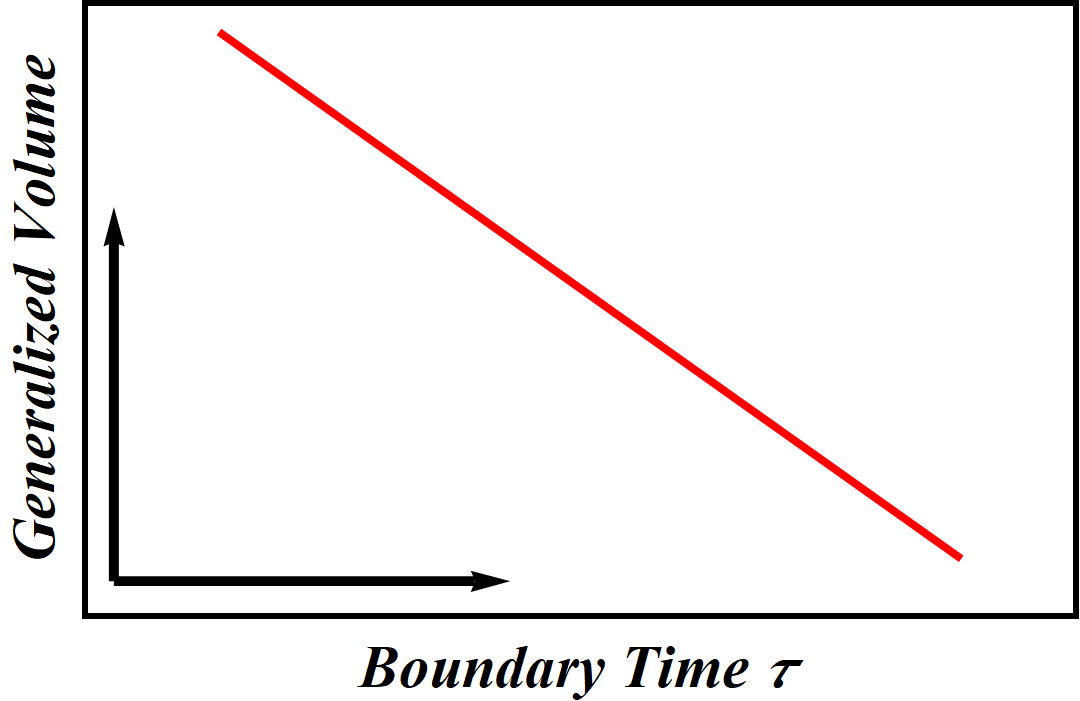}
		\caption{The surface close to the singularity.}
		\label{CtauWb}
	\end{subfigure}
	\caption[123]{The relation between the boundary time $\tau$ and the generalized volume-complexity $\mathcal{C}$ for the case of $F_{1} \neq F_{2}$ with $L=1$, $m=1$, $\alpha=0.1$, $\lambda_{GB}=0.3$, $\lambda_{W_{1}}=0.5 \times 10^{-2}$ and $\lambda_{W_{2}}=0.5 \times 10^{-7}$. 
	The blue and red curves correspond to two extreme hypersurfaces that can evolve to the late time. 
	When $\tau \to \infty$, $r_{min}$ is $r_{f_{R}}$ and $r_{f_{L}}$, respectively. }
	\label{CtauW}
\end{figure}
The reason for this is that when $a(r)^{2}$ is replaced by $a_{1}(r)a_{2}(r)$, the integrand of $\mathcal{C}$ is no longer positive inside the event horizon. 
So we need to impose certain restrictions on $a(r)$. 
The modified $a(r)$ will ensure that the integrand of $\mathcal{C}$ is always positive inside the event horizon, i.e.
\begin{align}
	a_{1}(r)\equiv& \sqrt{a^{2}_{GB}(r)}=|1+\lambda_{GB} \alpha^{2} \mathcal{R}_{GB}| ,\notag\\
	a_{2}(r)\equiv& \sqrt{a^{2}_{W}(r)}=|1+\lambda_{W_{1}}L^{4}C^{2}-\lambda_{W_{2}}L^{8}C^{4}|.
\end{align}
where the parameters and scalar functionals we choose are consistent with those in Subsec.~\ref{secGB}.
In this premise, if we order $U_{1}(r)=-f(r)a_{1}^{2}(r)r^{2(d-1)}$, Eq.~\eqref{C12} can be rewritten as 
\begin{equation}
	\mathcal{C}_{F_{1},\Sigma_{F_{2}}}=-\frac{2V_{d-1}}{G_{N}L}\int_{r_{min}}^{\infty}\frac{\sqrt{U_{1}(r)U_{2}(r)}}{f(r)\sqrt{P_{v}^{2}-U_{2}(r)}}dr.
\end{equation}
This conclusion is consistent with what Ref.~\cite{Belin:2021bga} discussed.
And it's growth rate is only analytically solved at a later stage, i.e.
\begin{equation}
	\lim_{\tau \to \infty} \frac{d\mathcal{C}_{F_{1},\Sigma_{F_{2}}}}{d \tau}=\frac{V_{d-1}}{G_{N}L}\sqrt{U_{1}(r_{f})}. 
\end{equation}
It is worth noting that when $r_{min} \to r_{f}$, $U_{2}(r)$ is at the local maximum, but the value of the growth rate is determined by $U_{1}(r)$ as shown in Fig.~\ref{U1rfC2}. 
\begin{figure}
	\centering
	\includegraphics[scale=0.2]{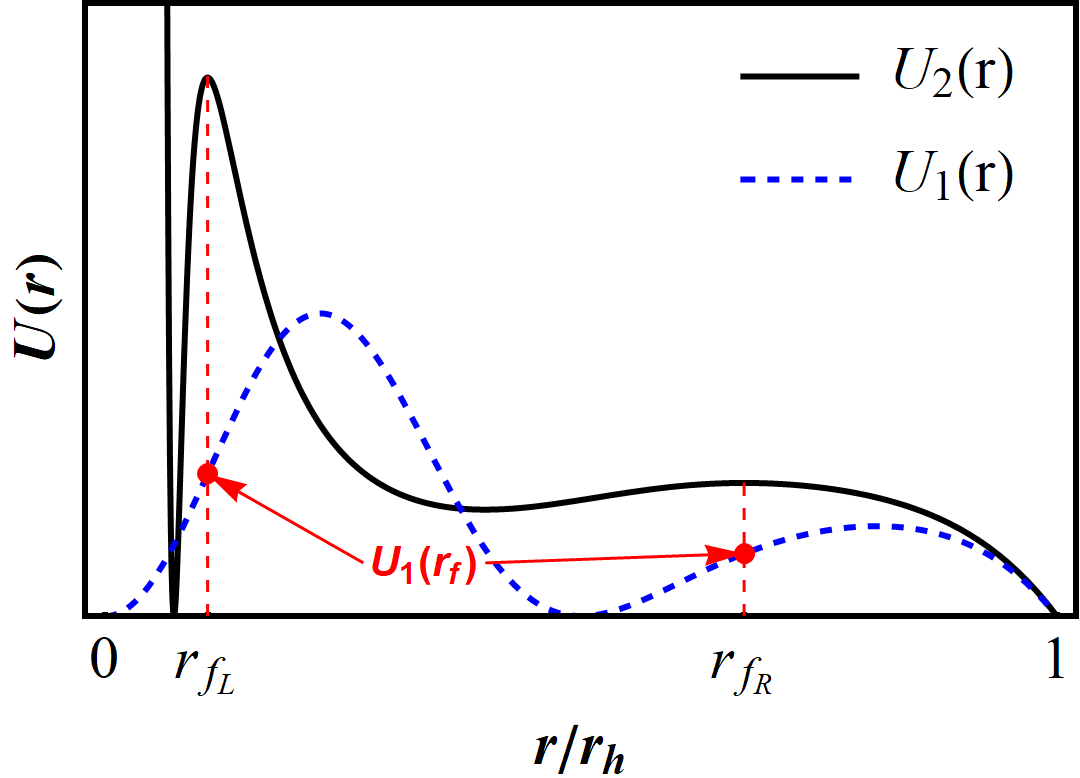}
	\caption[123]{The shape of the effective potential for the case of $F_{1} \neq F_{2}$ with $L=1,~m=1,~\alpha=0.01,~\lambda_{GB}=0.3,~\lambda_{W_{1}}=0.3\times 10^{-3},~\lambda_{W_{2}}=0.3\times 10^{-10} $}
	\label{U1rfC2}
\end{figure}
We can still find the turning time in this situation, as shown in the right figure of Fig.~\ref{U1rfC2turning}. 
\begin{figure}[htbp]
	\centering
	\begin{subfigure}[b]{0.45\textwidth}
		\centering
		\includegraphics[scale=0.17]{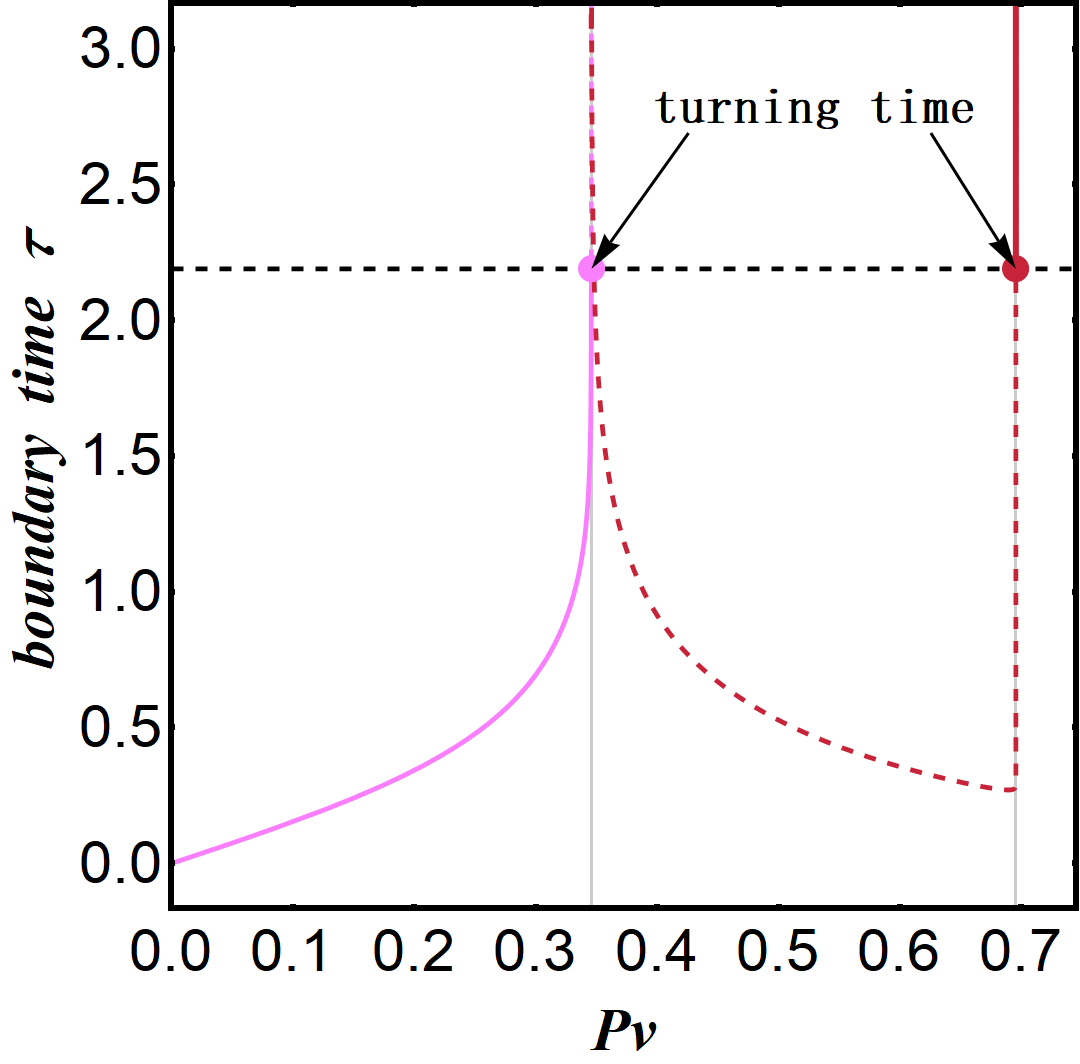}
		\caption{Boundary time $\tau$.}
		\label{U1rfC2turninga}
	\end{subfigure}
    \hspace{0.05\textwidth}
	\begin{subfigure}[b]{0.45\textwidth}
		\centering
		\includegraphics[scale=0.17]{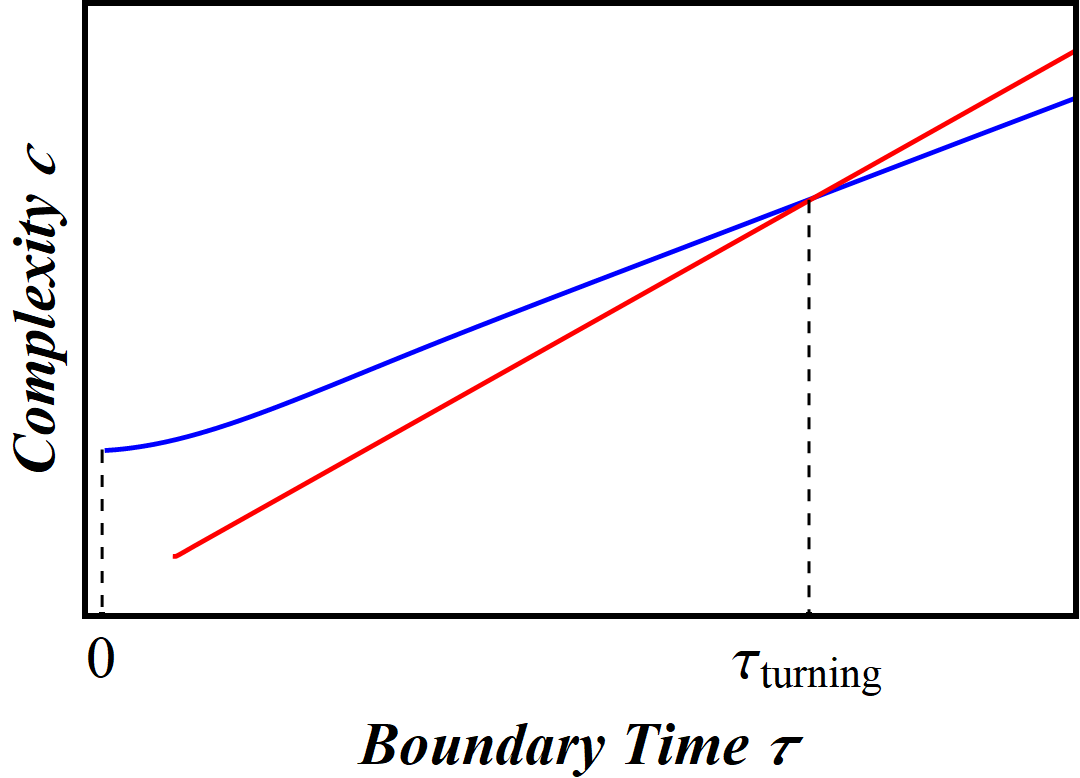}
		\caption{Generalized volume growth}
		\label{U1rfC2turningb}
	\end{subfigure}
	\caption[123]{The figure of the growth of generalized volume-complexity with time and the turning time for the case of $F_{1} \neq F_{2}$ with 
	$L=1,~m=1,~\alpha=0.01,~\lambda_{GB}=0.3,~\lambda_{W_{1}}=0.3\times 10^{-3},~\lambda_{W_{2}}=0.3\times 10^{-10} $.
	(a): The relation between the boundary time $\tau$ and the conserved momentum $P_{v}$.
	(b): Generalized volume-complexity evolution over time.}
	\label{U1rfC2turning}
\end{figure}
Naturally, we can also exchange $a_{1}(r)$ and $a_{2}(r)$, i.e., 
\begin{align}
	a_{1}(r)\equiv& \sqrt{a^{2}_{W}(r)}=|1+\lambda_{W_{1}}L^{4}C^{2}-\lambda_{W_{2}}L^{8}C^{4}|,\label{a1C2C4}\\
	a_{2}(r)\equiv& \sqrt{a^{2}_{GB}(r)}=|1+\lambda_{GB} \alpha^{2} \mathcal{R}_{GB}|\label{a2GB},
\end{align}
and can still obtain the turning time as shown in Fig.~\ref{U1rfGB}.
\begin{figure}[htbp]
	\centering
	\begin{subfigure}[b]{0.45\textwidth}
		\centering
		\includegraphics[scale=0.18]{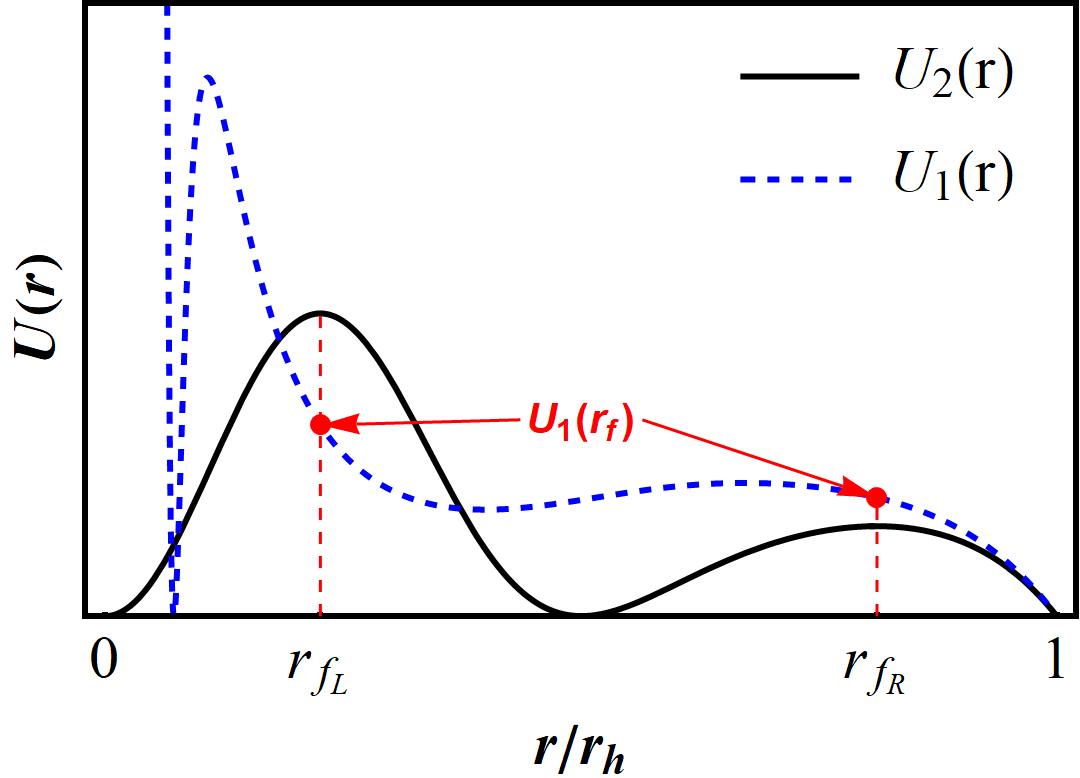}
		\caption{Effective potential $U(r)$.}
		\label{U1rfGBa}
	\end{subfigure}
    \hspace{0.05\textwidth}
	\begin{subfigure}[b]{0.45\textwidth}
		\centering
		\includegraphics[scale=0.17]{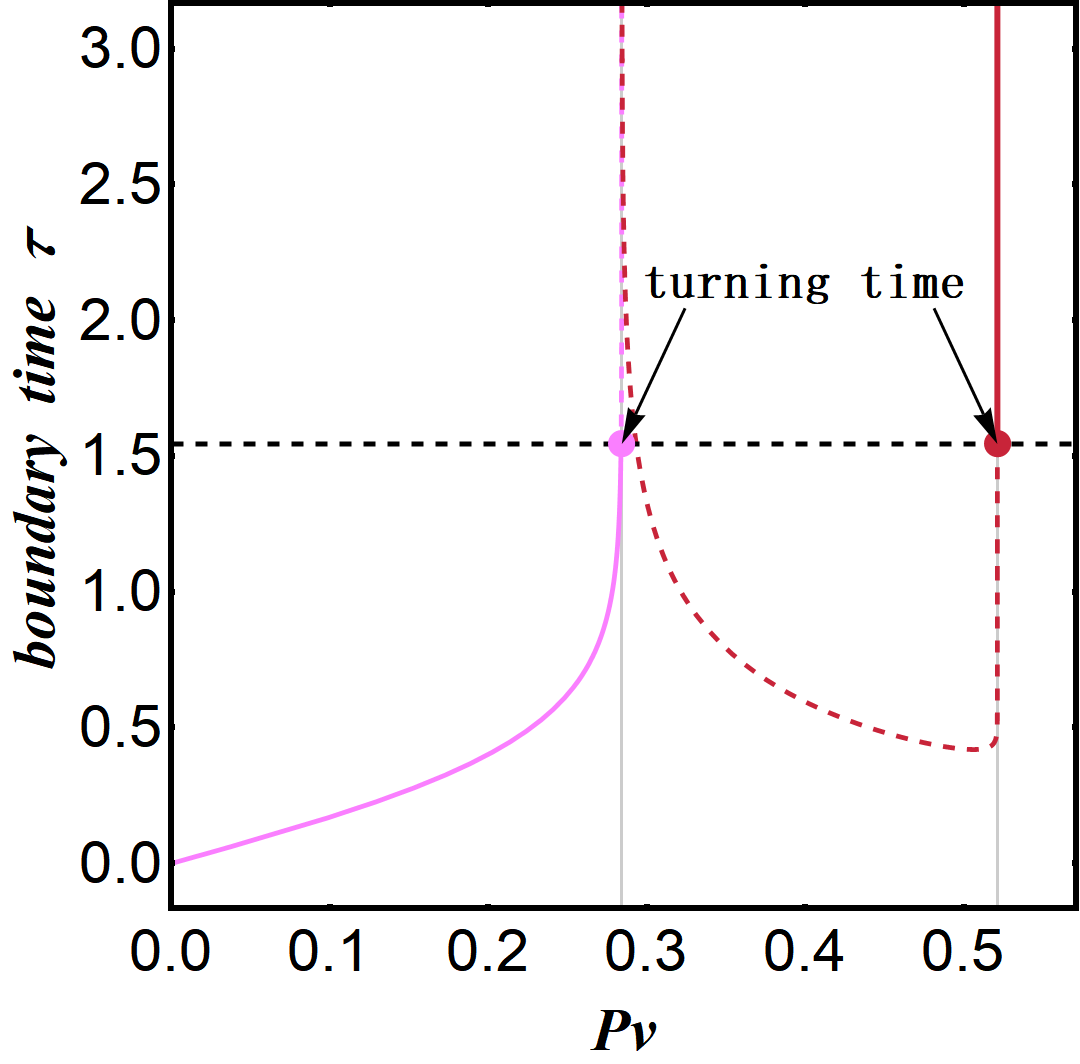}
		\caption{Boundary time $\tau$.}
		\label{U1rfGBb}
	\end{subfigure}
	\caption[123]{The evolution of the generalized volume-complexity over time after swapping $a_{1}(r)$ and $a_{2}(r)$ in Fig.~\ref{U1rfC2}. 
		(a): The effective potential for the case of $F_{1} \neq F_{2}$ with $L=1,~m=1,~\alpha=0.01,~\lambda_{GB}=0.3,~\lambda_{W_{1}}=0.3\times 10^{-3},~\lambda_{W_{2}}=0.3\times 10^{-10} $.
		(b): The relation between the boundary time $\tau$ and the conserved momentum $P_{v}$. 
		}
	\label{U1rfGB}
\end{figure}
\subsection{Differences from $F_{1}=F_{2}$}
For the case of $F_{1} \neq F_{2}$, since the generalized volume depends on the two different scalar functions, $a_1(r)$ and $a_2(r)$, 
a higher peak no longer necessarily corresponds to an extremal hypersurface with a larger generalized volume as shown in Fig.~\ref{U1rfSL}, 
where we follow the scalar function defined by Eqs.~\eqref{a1C2C4} and \eqref{a2GB}. 
For this case, the generalized volume corresponding to the left peak will never exceed that of the right, even if it is higher. 
\begin{figure}[htbp]
	\centering
	\begin{subfigure}[b]{0.45\textwidth}
		\centering
		\includegraphics[scale=0.17]{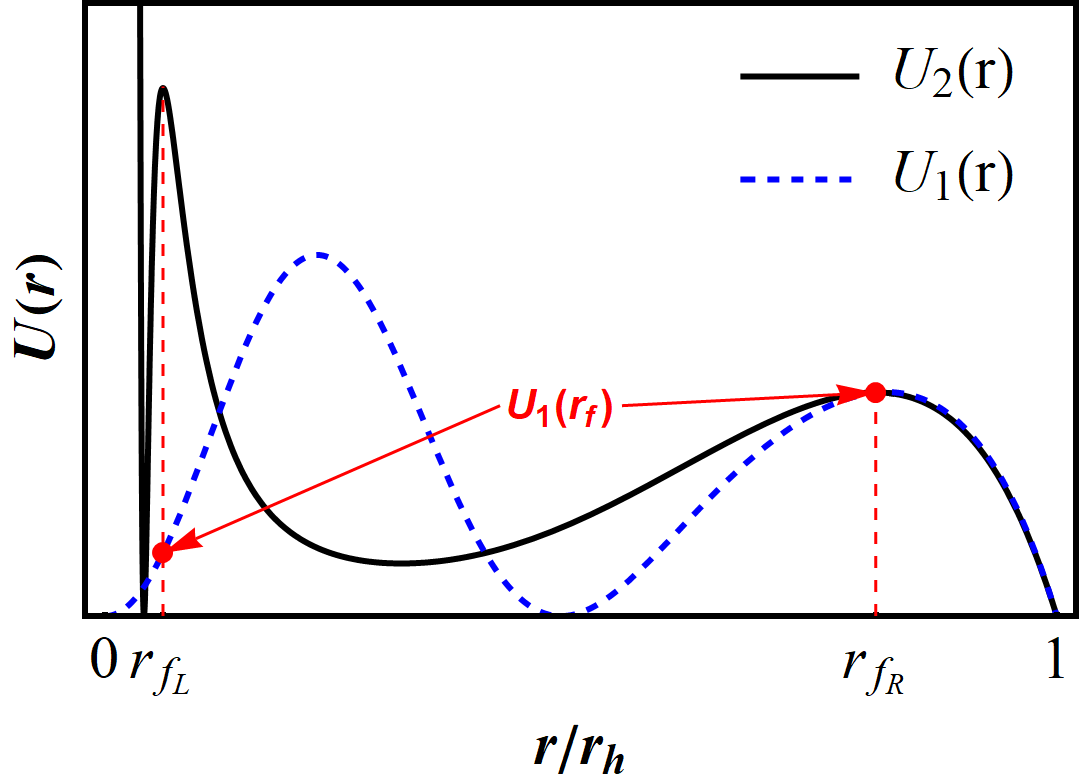}
		\caption{Effective potential $U(r)$.}
		\label{U1rfSLa}
	\end{subfigure}
    \hspace{0.05\textwidth}
	\begin{subfigure}[b]{0.45\textwidth}
		\centering
		\includegraphics[scale=0.17]{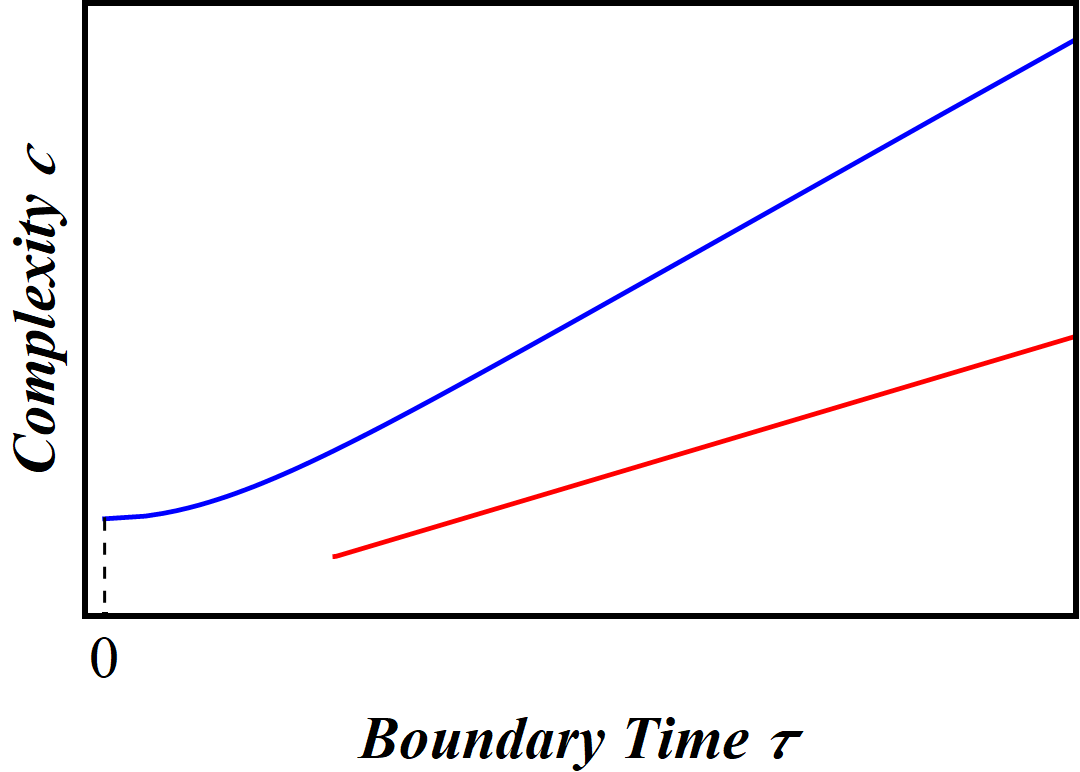}
		\caption{Generalized volume growth}
		\label{U1rfSLb}
	\end{subfigure}
	\caption[123]{
		(a): The effective potential for the case of $F_{1} \neq F_{2}$ with $L=1,~m=1,~\alpha=0.01,~\lambda_{GB}=1,~\lambda_{W_{1}}=10^{-4},~\lambda_{W_{2}}=10^{-12} $.
		(b): Generalized volume-complexity evolution over time with the above parameters.
		}
	\label{U1rfSL}
\end{figure}
On the other hand, unlike $F_1=F_2$, the evolution of the generalized volume-complexity over time may not be monotonous as shown in Fig.~\ref{Cgendecrease}; 
\begin{figure}[htbp]
	\centering
	\begin{subfigure}[b]{0.45\textwidth}
		\centering
		\includegraphics[scale=0.17]{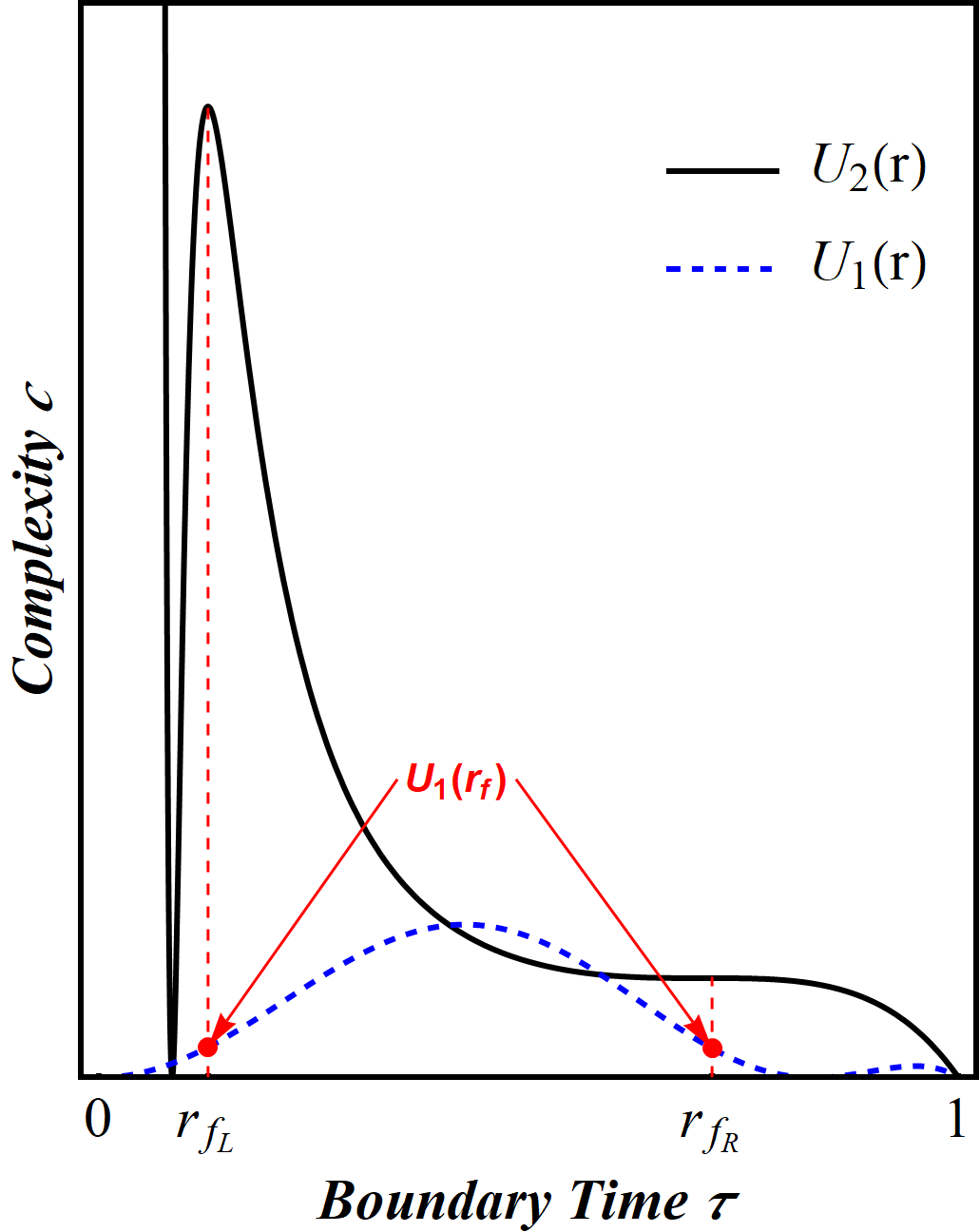}
		\caption{Effective potential $U(r)$.}
		\label{Cgendecreasea}
	\end{subfigure}
    \hspace{0.05\textwidth}
	\begin{subfigure}[b]{0.45\textwidth}
		\centering
		\includegraphics[scale=0.17]{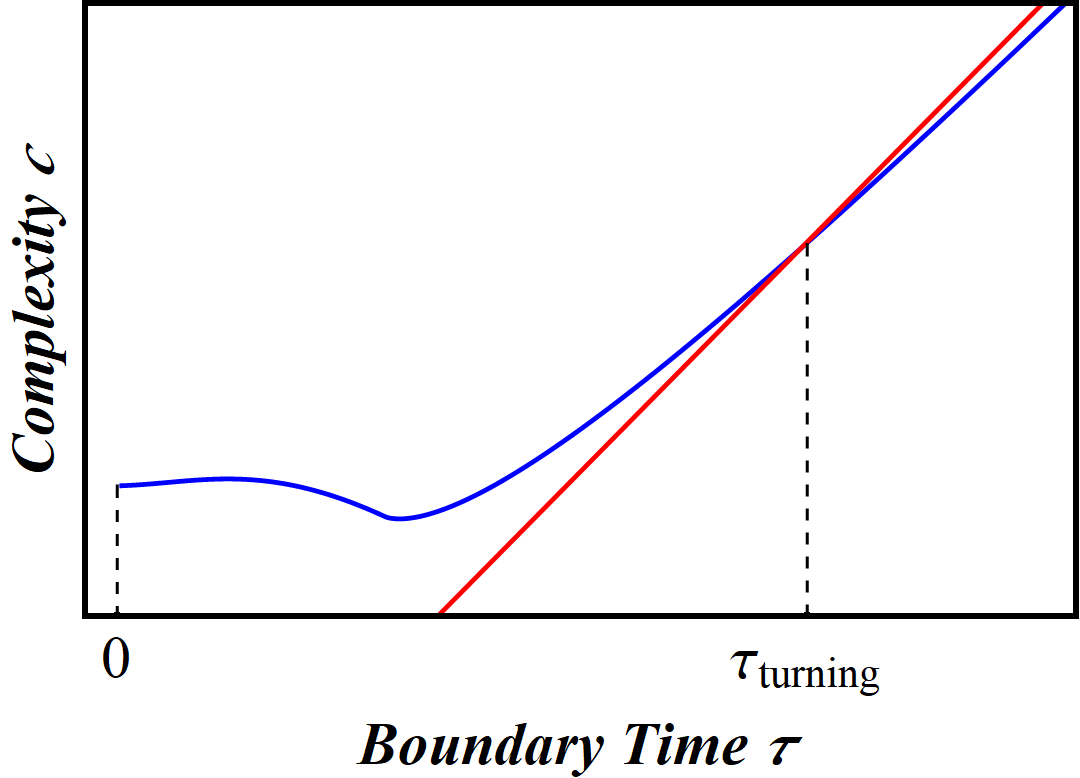}
		\caption{Generalized volume growth.}
		\label{Cgendecreaseb}
	\end{subfigure}
	\caption[123]{
		(a): The effective potential for the case of $F_{1} \neq F_{2}$ with $L=1,~m=1,~\alpha=0.1,~\lambda_{GB}=0.3,~\lambda_{W_{1}}=0.7 \times 10^{-2},~\lambda_{W_{2}}=10^{-8} $.
		(b): Generalized volume-complexity evolution over time with the above parameters.}
	\label{Cgendecrease}
\end{figure}
this does not affect the linear growth at late time. }

\section{Discussion} \label{sec6}
In this paper, we considered the generalized volume-complexity for the planar AdS black hole, the
five-dimensional Gauss-Bonnet-AdS black hole, the charged BTZ black hole, and rediscussed the four-dimensional RN-AdS black hole. 
We derived the expressions for the complexity, Eq.~\eqref{Cdr}, and the growth rate of complexity, Eq.~\eqref{CPv}.
The growth rate of complexity is directly proportional to the conserved momentum $P_{v}$.
We used some different scalar functions $a(r)$ to construct various gravitational observables. 
Through this process, we discovered that the effective potentials exhibit diverse behaviors depending on the choice of $a(r)$,
which corresponds to different behaviors in the growth rate of complexity.

In the case we analyzed, there are two or more local maxima for the effective potential.
Through numerical integration to solve inversely the boundary times with Eq.~\eqref{tau},
we observed that in the $P_{v}-\tau$ diagram, multiple $P_{v}$ values can intersect at the same boundary time $\tau$.
It means there are multiple extremal hypersurfaces that anchor to the same boundary time.
We chose the one with the largest generalized volume as the dual of the complexity.
However, the relative sizes of the generalized volumes are not fixed, the smaller one would ``surpass" at a certain moment.
We refer to this moment as the turning time. 
It is worth mentioning that the existence of the turning time does not depend on the space-time background inside the black hole, but can always exist by choosing different gravitational observables. 
Additionally, the turning time can occur more than once. 
Furthermore, we found that the turning time has a logarithmic relationship with the parameters.


In addition, we divided the generalized volume-complexity into
two categories based on their properties at the singularity or the Cauchy horizon.
Different categories determine whether the parameter space is contained within the full parameter space. 
Further, by taking the four-dimensional RN-AdS black hole as an example, we find that for black holes with two horizons, the growth rate of the generalized volume-complexity at late time may change discontinuously as the Cauchy horizon disappears. 

{On the other hand, we find that for the case of $F_{1} \neq F_{2}$, in order to satisfy the property of late-time linear growth of complexity, 
some restrictions need to be imposed for the scalar function $a(r)$ to ensure that the integrand of the generalized volume-complexity remains positive. 
On this basis, the turning time can still exists, the difference lies in the fact that during the relatively early stages, the evolution of complexity with time may not necessarily be monotonic.}

Finally, it should be pointed out that we can also consider the complexity equals
anything conjecture under the quantum correction, just like the quantum BTZ black hole.

\section*{Acknowledgments}
We would like to thank Le-Chen Qu and Shan-Ming Ruan for their very useful comments and discussions.
We also thank Jing Chen, Yi-Yang Guo, Jun-Jie Wan, Shan-Ping Wu and Yuan-Ming Zhu for their selfless help.
This work was supported by National Key Research and Development Program of China (Grant No. 2021YFC2203003), 
the National Natural Science Foundation of China (Grants No. 11875151 and No. 12247101), 
the 111Project under (Grant No. B20063), the Major Science and Technology Projects of Gansu Province, and Lanzhou City's scientic research funding subsidy to Lanzhou University. 

\appendix
\section{JT black holes}\label{appa}
\begin{figure}[htbp]
    \centering
    \includegraphics[scale=0.25]{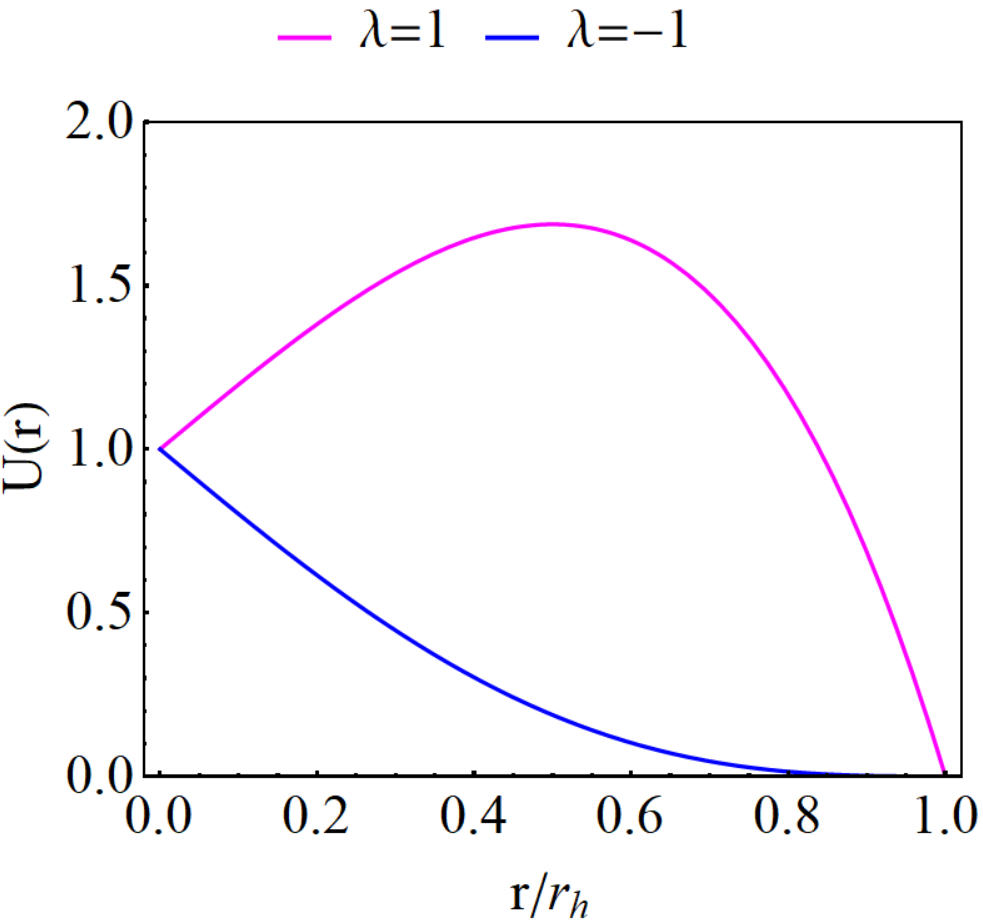}
    \caption[123]{ The shape of the effective potential for $L = \phi_{h} = \phi_{0} = 1$. The purple line has one peak, and
	the bule line has no peak.}
	\label{JTUr}
\end{figure}

In this appendix, we will consider briefly about the JT black holes, whose matrix is
\begin{align}
	ds^{2}&=-f(r)dt^{2}+\frac{1}{f(r)}dr^{2},\notag\\
	f(r)&=\frac{r^{2}}{L^{2}}\left(1-\frac{r_{h}^{2}}{r^{2}}\right).
\end{align}
It is natural that we can use the dilaton field to construct scalar functions, that is
\begin{equation}
	a(r)=1+\lambda\frac{\phi(r)}{\phi_{0}}
\end{equation}
where
\begin{equation}
	\phi(r)=\phi_{h}\frac{r}{r_{h}},
\end{equation}
$\phi_{0}$ is a proportionality factor in place of the Newton's constant, and the integration constant
$\phi_{H}$ is the value of $\phi$ on the horizon. In this case,
\begin{equation}
	U(r)=-f(r)a^{2}(r),
\end{equation}
we can get a shape of the effective potential from this, as shown in Fig.~\ref{JTUr}.
The JT black hole does not have a singularity, and the $f(r)|_{r=0}=r_{h}^{2}/L^{2}$ is a non-zero
constant, it results in the effective potential's starting point not being zero or infinity.
The characteristics of the effective potential at $r = 0$ depends only on the choice of the
scalar functions. However, we think it can be contained by the first category. Whether the
effective potential diverges at zero does not by itself affect the extent of the phase space or
the size of the local maximum.

 ~\\
 ~\\
 ~\\

\end{document}